\documentstyle[aaspp4]{article}

\received{1998 July 3}
\accepted{1998 October 13}

\slugcomment{To appear in ApJ}

\lefthead{Richards et al.}
\righthead{Intrinsic \ion{C}{4} Absorption}

\begin{document}

\title{Determining the Fraction of Intrinsic \ion{C}{4} Absorption in Quasi-Stellar Object Absorption Line Systems}

\author{Gordon T. Richards\altaffilmark{1} and Donald G. York\altaffilmark{2}}
\affil{Department of Astronomy and Astrophysics, University of Chicago, 5640 S. Ellis Avenue, Chicago, IL 60637}
\authoremail{richards,don@oddjob.uchicago.edu}

\author{Brian Yanny and Ronald I. Kollgaard }
\affil{Fermi National Accelerator Laboratory, Kirk Road and Pine Street, Batavia, IL 60510}
\authoremail{yanny,rik@sdss.fnal.gov}

\author{S. A. Laurent-Muehleisen}
\affil{UC-Davis and IGPP/LLNL, L-413, 7000 East Ave., Livermore, CA 94550}
\authoremail{slauren@igpp.llnl.gov}

\and

\author{Daniel E. Vanden Berk}
\affil{McDonald Observatory and Department of Astronomy, University of Texas at Austin, Austin, TX 78712}
\authoremail{danvb@astro.as.utexas.edu}

\altaffiltext{1}{Also, Fermi National Accelerator Laboratory, Kirk Road and Pine Street, Batavia, IL 60510} 
\altaffiltext{2}{Also, Enrico Fermi Institute, 5640. S. Ellis Avenue, Chicago, IL 60637} 

\begin{abstract}

We present the results of an exhaustive study of QSO Absorption Line
Systems (QSOALSs) with respect to intrinsic QSO properties using an
updated catalog of data in the literature.  We have searched the
literature for 6 and $20\,{\rm cm}$ radio flux densities and have
studied $20\,{\rm cm}$ contour plots from the Faint Images of the
Radio Sky at Twenty centimeters (FIRST) VLA Survey in order to compare
the absorption properties with radio luminosity, radio spectral index
and radio morphology.  Although the data in our catalog are decidedly
heterogeneous, great care has been taken to account for potential
biases during the course of our research.  Analysis of relatively
unbiased subcatalogs allows us to investigate the properties of
QSOALSs with better statistics than with any single homogeneous
catalog.  This work focuses particularly on the nature of \ion{C}{4}
QSOALSs and their distribution in velocity space in light of intrinsic
QSO properties.  We find that the distribution of narrow, \ion{C}{4}
absorption systems with relative velocities exceeding $5000\,{\rm
km\,s^{-1}}$ is dependent not only on the optical luminosity of the
QSOs, but also on the radio luminosity, the radio spectral index and
the radio morphology of the QSOs.  These observations are apparently
inconsistent with the hypothesis that these systems are predominantly
due to intervening galaxies and it would seem that the contamination
of the intervening systems (from 5000 to $75,000\,{\rm km\,s^{-1}}$)
by those that are intrinsic to the environment of the QSO is
significantly larger than expected.  We stress the need for truly
homogeneous and unbiased surveys of QSOALS to confirm these results
from our inhomogeneous data set.
\end{abstract}


\section{Introduction}

Although there was great debate (e.g., \cite{bs66}) over the origin of
QSO absorption line systems for many years after they were first
discovered (\cite{blb66}; \cite{sl66}), current interpretation treats
systems with $v>5000\,{\rm km\,s^{-1}}$ as arising from intervening
galaxies.\footnote[3]{Throughout this paper velocities ($v$) refer to
the blueshifted velocity relative to the emission-line redshift of the
QSOs, where $\beta(=v/c)$ is given by eq.(1) in \S 3.2.}  Time and
time again, observations have shown that narrow, \ion{C}{4} absorbers
with velocities exceeding $5000\,{\rm km\,s^{-1}}$ are distributed in
a way that is consistent with the absorbers being caused by
intervening galaxies (\cite{ysb82}; \cite{sbs88}; \cite{ste90}).  It
has also been shown that \ion{C}{4} absorption systems within
$5000\,{\rm km\,s^{-1}}$ of the QSO are distinctly different from
those at higher relative velocities and are probably intrinsic to the
QSO environment; these are primarily strong systems and are biased
towards steep-spectrum radio-loud objects (\cite{fwp+86};
\cite{awf+87}; \cite{fcw+88}).  Equally well accepted is that troughs,
observed to at least $0.1{\rm c}$, in the spectra of so-called Broad
Absorption Line QSOs (BALQSOs) must be intrinsic to the QSO
(\cite{tur88}).  However, even though QSOs can clearly accelerate
material to velocities exceeding $5000\,{\rm km\,s^{-1}}$, few have
argued that any significant fraction of the narrow, \ion{C}{4} lines
blueward of $5000\,{\rm km\,s^{-1}}$ are related to the QSO.

In spite of the above statements, a search of the literature would
turn up any number of examples of narrow, high ionization QSOALS that
are intrinsic absorption candidates.  Perry, Burbidge, \& Burbidge
(1978) present evidence that narrow, high velocity absorption lines
may be intrinsic to OQ\,172 and PHL\,957 since their fine structure
lines would seem to indicate very large electron densities.  Weymann
et al. (1979) discuss evidence for an ejected component of narrow
absorption systems out to velocities of $18,000\,{\rm km\,s^{-1}}$.
Young et al. (1982) claim to see an excess of relatively narrow
systems with $\beta<0.1\,{\rm c}$ in 3 QSOs.  Even though these are
BALQSOs, it establishes the possibility of {\em narrow}, high velocity
systems.  Foltz et al. (1986) find that the excess of associated
absorption is dominated by strong systems and that there are 3 such
systems in their sample with velocities greater than $5000\,{\rm
km\,s^{-1}}$.  More recently, Borgeest \& Mehlert (1993) concluded
that the ``association hypothesis'' was more probable than the
``intervening hypothesis'' for a large fraction of the \ion{C}{4}
systems in 5 surveys of QSO absorption lines.  Petitjean, Rauch, \&
Carswell (1994) have found super-solar abundances in 6 systems with $v
< 13,000\,{\rm km\,s^{-1}}$ and postulate that these systems are
intrinsic.  Arav, Li, \& Begelman (1994) have modeled BALQSOs and find
that the terminal velocity of radiatively accelerated clouds is on the
order of 16,000 to $25,000\,{\rm km\,s^{-1}}$.  Barlow, Hamann, \&
Sargent (1997) acknowledge the existence of narrow, intrinsic systems
and give nine ways to tell the difference between a system that is
intrinsic and one that is intervening.  Jannuzi et al. (1996) describe
a broad \ion{C}{4} absorber at $56,000\,{\rm km\,s^{-1}}$ in PG
2302+039.  Hamann et al. (1997) find evidence for relatively narrow,
intrinsic absorption from $1500$ to $51,000\,{\rm km\,s^{-1}}$ in PG
0935+417, UM675 and Q2343+125.

In light of the above evidence, circumstantial as it may be, it should
come as no surprise if it were found that a significant fraction of
narrow \ion{C}{4} lines with high velocities turned out to be ejected
material from QSOs rather than intervening galaxies.  In this work we
consider all \ion{C}{4} systems whose absorption redshifts are such
that they are within $71,400\,{\rm km\,s^{-1}}\,(\beta=v/c=0.238)$, at
which point \ion{C}{4} systems become blended with the Lyman $\alpha$
forest.

QSO absorption line systems may originate from any or all of 4
distinct environments including (1) the host galaxy of the QSO, (2)
material ejected by the QSO, (3) the cluster environment of the QSO
and (4) intervening galaxies along the line of sight.  Throughout this
paper we refer to any absorber that could be strongly affected by the
radiation field of the QSO (which includes the first 3 classes listed
above) as intrinsic.  Note that this is in contrast to the definition
that some have used in other work.

This body of work utilizes the QSO absorption line data collected by
York et al. (1991), and began as a result of previous papers that draw
upon the data from this catalog.  Specifically, Vanden Berk et
al. (1996) found that QSOs that have more \ion{C}{4} systems along
their line of sight tend to be brighter than those that have fewer
such systems.  Gravitational lensing by intervening galaxies was
postulated as the origin of the effect, but intrinsic absorption could
not be ruled out.  The gravitational lensing hypothesis was
investigated by Holz \& Wald (1998), whereas the present work
concentrates on the possibility of intrinsic absorption.

As such, the purpose of this work is to attempt a preliminary
quantification of the fraction of narrow, but intrinsic \ion{C}{4}
absorption in QSOs.  If this fraction is anything other than
negligible, then studies of absorption line systems will have to be
reconsidered.  In addition, we comment briefly on the properties of
$z_{abs} \approx z_{em}$ absorbers, line-locking, and the effects of
gravitational lensing on our analysis.  Section 2 briefly describes
the data.  In \S 3, we present an analysis of the \ion{C}{4}
distribution in velocity space with respect to various QSO properties.
In \S 4 we discuss the implications of our analysis.  Section 5
presents our conclusions.  In the Appendix, we present a more detailed
description of the data.

\section {The Data}

The data gathered by York et al. (1991) has been updated to include
most of the absorption line data in the literature up to 1994 October
and some of the more recent data up to 1997.  Since the catalog is
inhomogeneous, care must be taken to avoid any possible biases in the
data.  In particular, we correct for known biases by ensuring that all
stated results apply for an analysis of only the largest possible
``homogeneous'' samples, which will be defined later.

Given the possibility of intrinsic absorption and coupling it with the
fact that radio-loud and radio-quiet QSOs could be separate
populations, one has the potential for some significant biases.
Therefore we have undertaken the task of compiling the radio
properties of all of the QSOs in the catalog.  We have searched the
literature for 6 and 20\,cm integrated radio flux densities for our
objects in order to classify the QSOs as radio-loud (RL) or
radio-quiet (RQ) and (for those that are RL) as flat spectrum or steep
spectrum.  Additionally we have analyzed 20\,cm data from the Faint
Images of the Radio Sky at Twenty centimeters (FIRST; \cite{bwh95})
and NRAO VLA Sky Survey (NVSS; \cite{ccg+98}) VLA surveys to determine
the 20\,cm radio morphology and core-to-lobe ratios of $\sim 1/4$ of
the RLQSOs in our database.  We then use this information to study the
velocity distribution of high-ionization \ion{C}{4} absorption
relative to intrinsic QSO properties (redshift, absolute visual
magnitude, radio spectral index, and radio morphology).

For a more detailed description of the absorption line data, we refer
the reader to York et al. (1991, and references therein).  The details
regarding the collection and analysis of the intrinsic QSO properties
is essential to this work; however, we feel that a discussion of it
here would detract from the focus of the paper, and we have chosen to
leave that discussion to the Appendix.

\section{Analysis and Results}

To ensure the reality of the \ion{C}{4} systems used in the analysis
of our samples, we have made some stringent cuts on the data.  First
of all, only systems classified as grade A, B, or C are used.  The
grading system for the new catalog is derived from that of York et
al. (1991), but has been revised to some extent.  Herein grades A, B,
and C refer to systems with four or more lines, three lines, and two
lines, respectively.  In addition, we require that both members of the
doublet have a rest equivalent width (REW) greater than $0.15{\rm
\AA}$ to avoid any problems with weak lines appearing preferentially
on top of emission lines due to higher signal-to-noise ratio (S/N;
\cite{sbs88}).  Note that this requirement means that grade C (only
two lines) systems are ``doublet only'' systems.  This constraint also
decreases the probability of the two stronger lines of line-locked
systems appearing to be a single system.  After all the above cuts
have been applied, systems within $250\,{\rm km\,s^{-1}}$ of each
other are combined (\cite{sbs88}).  The new redshift is the equivalent
width weighted mean of the redshifts of the two systems being merged.
If both systems are from the same reference then the rest equivalent
widths are summed, and the errors are added in quadrature.  If the
absorbers are not from the same reference then the REW is given by the
error weighted mean of the individual REWs.  Finally, all lines are
required to be at least $5\,\sigma$ detections.

Then, for each line meeting the above requirements, we determine how
many times the absorber could have been seen at this velocity (with
this strength) for all of the QSOs in the catalog.  This number is
then used to normalize the counts, since the bins are not necessarily
sampled evenly.  See Vanden Berk et al. (1996) for specifics on this
normalization process.

The first step of our analysis was to look for systematic differences
(between different QSO properties) of the absorber distribution in
velocity space ($\beta$) relative to the QSOs.  For example, if we can
assume that the radio spectral indices of the QSOs are unlikely to be
affected by the presence of intervening galaxies it would seem
reasonable to guess that any systematic trends between radio spectral
index and absorber velocity distributions are due to absorption that
is intrinsic to the QSO.  These systematic differences could be
manifested as a change in the normalized number of absorbers per unit
velocity difference with respect to the QSO redshift ($d{\cal
N}/d\beta$), as a perturbation in a small velocity range as might be
caused by line-locking effects, or as a large scale difference in the
shapes of the velocity distribution.

Each QSO was placed into one class of each of the following sets of
categories: (1) bright or faint, (2) high redshift or low redshift,
(3) radio loud, radio moderate, or radio quiet, (4) flat spectrum or
steep spectrum, and (5) core dominated or lobe dominated.  The
bright/faint division was made by determining the median absolute
magnitude of the sample, which is $M_V = -27.0$.  High redshift versus
low redshift was also found by taking the median of the sample which
is $z_{em} = 2.0$.  Separation into radio luminosity bins is discussed
in the Appendix.  For those QSOs with both 6 and 20\,cm flux densities
we find a median radio spectral index of $\alpha_6^{20} = -0.5$, and
we take QSOs with $\alpha < -0.5$ to be steep-spectrum.
(N.B. throughout this paper spectral indices ($\alpha$) are given such
that $f_{\nu} \propto \nu^{\alpha}$.)  Radio morphology was determined
by eye on a case by case basis from FIRST contour plots that were
kindly provided by R. Becker and also by determining the core-to-lobe
ratios using both FIRST and NVSS data from the VLA.  See the Appendix
for further details regarding the QSO properties.

In addition to comparing all of the absorbers in a single population
to all absorbers in another population, we have also done the
comparisons after attempting to normalize for other effects.  For
example we find that the brighter QSOs are statistically at higher
redshifts than the fainter QSOs; so how can we be sure that an effect
seen in the bright versus faint distribution is not really due to
redshift?  Obviously we cannot be sure that the redshift is not the
cause of the effect unless we have some way of normalizing the
redshift distributions between the bright and faint samples.

In order to handle biases of the type described above, we have devised
a naive, yet effective scheme to minimize these effects.  Very simply,
we force the distribution of a given parameter or parameters to be the
same for the two samples we want to compare.  This is done by binning
the data in the parameter that we want to normalize out and requiring
that the same number of each type of QSO appear in each bin.  In a
given bin, we keep all of the QSOs of the type with the least number
in that bin and we randomly select from the complementary QSO
population as many QSOs as matches the lesser number.  We then make 10
such random catalogs for each pair of properties that we wish to
compare.  For example, if there are eight bright QSOs in the $2.0 <
z_{em} < 2.5$ bin and only five faint QSOs, then our subcatalog will
have all five of the faint QSOs and five randomly selected bright QSOs
in this redshift bin.  We repeat this 10 times, making 10 subcatalogs
--- each of which are analyzed separately, then the individual results
are averaged.  The hope is that this will remove any effects of
parameters that are not being studied in a given sample.  We have
every indication that this is working as intended and therefore are
comfortable that the reported results are indeed due to the properties
in question and not a result of correlations between intrinsic QSO
properties.

\subsection{Overview of the Large Scale Velocity Distribution of \ion{C}{4}}

Previous work (e.g., \cite{ysb82}; \cite{yc94}) has stressed that the
distribution of \ion{C}{4} absorbers is uniform in velocity space (at
least out to $\beta = 0.238$ where the Lyman $\alpha$ forest makes it
difficult to properly identify \ion{C}{4} absorption).  This is
generally done by showing that the Bahcall-Peebles parameter
(\cite{bp69}) is distributed smoothly, or by using a
Kolmogorov-Smirnov (K-S) test to show that the velocity distribution
of absorbers is consistent with a flat distribution.  In our analysis
we chose to do neither of the above.  We feel that the data are too
inhomogeneous for either of the Bahcall-Peebles tests (though ``Test
2'' was designed for inhomogeneous data).  As for the K-S test, there
is no need to repeat what has already been done numerous times and
more importantly we do not believe that doing such a test on our data
would yield any new results.  That is, we will not argue that the
distributions are inconsistent with being flat.

\subsection{Bright versus Faint}

The distribution of \ion{C}{4} absorption in velocity space with
respect to absolute optical magnitude can be seen in
Figure~\ref{fig1}, where we plot $d{\cal N}/d\beta [\equiv N(\beta)]$
versus $\beta$ for intrinsically bright QSOs ({\em solid line}) and faint QSOs
({\em dashed line}).  Throughout this paper, $\beta$ is given by
\begin{equation}
\beta = \frac{\left(1+z_{em}\right)^2 - \left(1+z_{abs}\right)^2}{\left(1+z_{em}\right)^2 + \left(1+z_{abs}\right)^2}.
\end{equation}
The total number of absorbers contributing to the normalized histogram
is noted for each population and Poisson error bars are given.  In
these plots All refers to the sample that includes A, B, and C
graded absorbers that have been combined within $250\,{\rm
km\,s^{-1}}$.  In the lower panel the ZR refers to the same
sample, but with the redshift (z) and radio luminosity (R) normalized
out as per the procedure described in in the beginning of \S 3.

From Figure~\ref{fig1} it is clear the there is an excess of absorbers
in bright QSOs over faint QSOs from $15,000\,{\rm km\,s^{-1}}$ to
$65,000\,{\rm km\,s^{-1}}$ ($0.05 < \beta < 0.2167$) for both samples.
The bottom panel shows that if the redshift and radio luminosity
distributions are normalized as described above, then the effect still
persists, but at smaller significance.  In particular, although the
higher redshift QSOs in our sample have been shown to be statistically
brighter, the bright/faint separation is still evident in the redshift
normalized sample.

It is also interesting to note that there is a small peak in the
number density of absorbers in bright QSOs near $\beta = 0.1c$.  This
is near the typical terminal velocity of BAL troughs (\cite{tur88});
however, we have excluded all BALQSOs and possible/probable BALQSOs
from our sample.  Also the peak is right where one would expect a
feature if there were significant line-locking between Si\,{\sc iv} and
\ion{C}{4} (\cite{bb75}).

Table~\ref{tbl-1} gives the significance of the differences in two
velocity bins for both of the bright/faint samples studied using a
simple Student's $t$-test (see the Appendix for details).  In
addition, we determine the significance for 2 other samples.  In
Table~\ref{tbl-1} 1000 B refers to the sample with A and B graded
absorbers that have been combined within $1000\,{\rm km\,s^{-1}}$,
whereas 1000 C has A, B, and C graded absorbers.  The 1000 C
sample is that which most closely resembles the primary sample used by
Sargent, Boksenberg, \& Steidel (1988).  In all, the bright/faint
difference is a $\ge 2.74 \sigma$ effect, which corresponds to a
$99.4\%$ confidence limit (assuming that the error distribution can be
considered approximately Gaussian).  At velocities $< 15,000\,{\rm
km\,s^{-1}}$ there is no apparent difference in the two populations,
and at higher velocities, the faint sample actually shows a small
excess.

\subsection{Loud versus Quiet}

The radio-loud and radio-quiet distributions (Fig.~\ref{fig2}) show
little difference at zero velocity, whereas there is a small, but
significant ($3.1\sigma$) excess in RQQSOs out to at least 0.1c.
Again it is interesting that the excess cuts off at about $0.1c$,
which is about the terminal velocity of broad absorption troughs, even
though we have removed the known BALs from our sample.  This excess
remains if the redshift (z) and absolute optical magnitude (M)
distributions are normalized for the two samples, which means that the
excess cannot be the result of a correlation with $z_{em}$ or $M_V$.
From Table~\ref{tbl-1}, we see that combining absorbers within
$1000\,{\rm km\,s^{-1}}$ also has little effect on the significance of
the excess of absorbers in RQ objects; however, rejection of the
``doublet only'' (grade C) systems results in no significant difference
between the two samples.  Nevertheless, there would appear to be a
difference in the distributions of absorbers in our primary sample ---
which poses problems for all the high velocity systems being caused by
intervening galaxies.

\subsection{High z versus Low z}

At first glance, the high redshift and low redshift samples
(Fig.~\ref{fig3}) also seem to show a difference in their mean
values of $d{\cal N}/d\beta$.  However, after normalization of the
$M_V$ distribution for the two samples, we find that there is no
significant difference between the two samples.  Therefore, any
apparent excess of absorbers along the line of sight to high redshift
QSOs is probably due to the tendency for the high $z$ QSOs to fall
into our bright sample.  (N.B. the bright/faint differences hold in
spite of redshift normalization, so the opposite statement is {\em
not} true.)  This lack of change in $d{\cal N}/d\beta$ for our two
different redshift samples is essentially the same as saying that
$d{\cal N}/dz$ is not a function of redshift, since there is no change
in the average level of $d{\cal N}/d\beta$ between our two redshift
samples.  However, other than the comments made in \S 4.3, we will not
comment further on the relationship between our analysis of $d{\cal
N}/d\beta$ and studies of the evolution of $d{\cal N}/dz$, which is
more commonly discussed in the literature.  For discussions of the
redshift evolution of \ion{C}{4}, see Vanden Berk et al. (1996),
Borgeest \& Mehlert (1993), or York et al. (1991).

\subsection{Steep versus Flat}

Perhaps the most striking results of our study come from the
comparison of steep-spectrum radio-loud QSOs to flat-spectrum
radio-loud QSOs (Figs.~\ref{fig4} and \ref{fig5}).  First, there is an
apparent excess of \ion{C}{4} absorbers at velocities less than
$5000\,{\rm km\,s^{-1}}$ ($\beta < 0.0167$) along the line of sight to
steep-spectrum sources --- an effect that was first reported by
Weymann et al. (1979) and confirmed by Foltz et al. (1986).
Essentially, Foltz et al. (1986) found that their excess and the lack
of an excess in the Young, Sargent, \& Boksenberg (1982) analysis for
$v<5000\,{\rm km\,s^{-1}}$ could be explained by the lack of
steep-spectrum QSOs in the YSB data set relative to that of Foltz et
al. (1986).  They also found that these ``associated'' systems tend to
be very strong (REW $> 1.5\,{\rm \AA}$).  This excess of associated
absorption is quite evident in Figure~\ref{fig6}, where we plot REW
versus $\beta$ for \ion{C}{4} absorption systems in both steep- and
flat-spectrum QSOs.  However, we caution that in Figure~\ref{fig6} the
number of absorbers is not normalized by the number of times that they
could have been seen.

Upon closer inspection of the top panel of Figure~\ref{fig4} we see
that the large scale ($10,000\,{\rm km\,s^{-1}}$ bins) distribution of
steep-spectrum absorbers near the QSO is really not significantly
greater than that of the flat-spectrum QSOs, but the apparent dearth
of higher-velocity absorbers in steep-spectrum QSOs makes the effect
appear more significant.  In fact, if we normalize the redshift (z),
absolute optical magnitude (M) and radio luminosity (R) distributions
for the flat and steep QSOs (as in Fig.~\ref{fig4}, {\em bottom}) then
we find that there is {\em no} difference between steep and flat for
$\beta c < 5000\,{\rm km\,s^{-1}}$.  However, on smaller scales
$(1500\,{\rm km\,s^{-1}})$ bins, we do confirm that excess in
steep-spectrum sources reported by Foltz et al. (1986).  This is an
interesting result and deserves further consideration, but it is
beyond the scope of this paper and we shall postpone any further
discussion.

For absorbers at velocities exceeding $5000\,{\rm km\,s^{-1}}$, both
distributions are relatively flat, which is consistent with the
intervening galaxy hypothesis; however, there is apparently a
significant difference in the mean value of $d{\cal N}/d\beta$.  In
particular, there is an excess of absorbers in flat-spectrum objects
over absorbers from steep-spectrum objects from $5000\,{\rm
km\,s^{-1}}$ to at least $25,000\,{\rm km\,s^{-1}}$ if not
$55,000\,{\rm km\,s^{-1}}$.  For the data that has been cut most
stringently, the difference between the average level of $d{\cal
N}/d\beta$ between flat and steep QSOs from $5000\,{\rm km\,s^{-1}}$
to $65,000\,{\rm km\,s^{-1}}$ is 2.84 and is a $4.2\sigma$ effect.
The largest excess is from $5000\,{\rm km\,s^{-1}}$ to {$25,000\,{\rm
km\,s^{-1}}$ for the sample which has been corrected for redshift (z),
optical absolute magnitude (M), and radio luminosity (R) effects.
Here there are seven more absorbers per unit $\beta$, per line of
sight in the flat-spectrum QSOs as compared to the steep-spectrum
QSOs.  As it is difficult to believe that the clouds causing the
absorption could significantly affect the radio spectral index, one is
left to conclude that these ``excess'' absorbers are likely to be
intrinsic to the QSO.

We have also determined that the QSO spectral indices correlate
roughly with radio morphology as is evidenced by comparisons with
FIRST $20\,{\rm cm}$ radio data.  This correlation lends credence to
the idea that the difference between the flat- and steep-spectrum
samples might be due to absorbers in the vicinity of the QSOs.  This
is important because variability is likely to have a deleterious
effect upon our spectral indices, but it is our hope that we have
enough data to dampen the effects of variability.  It is particularly
likely that the flat-spectrum QSOs will be included in the
steep-spectrum sample since the orientation of the flat-spectrum QSOs
is such that they are more likely to be beamed and thus variable.
This hypothesis is supported by our calculations of core-to-lobe
ratios (see the Appendix) where we find that for 35 QSOs with both
spectral indices and core-to-lobe ratios, there are 10 core-dominated
sources that we have classified as ``steep'', but only three
lobe-dominated sources that were classified as ``flat''.  A full
analysis is beyond the scope of this paper, but we will revisit the
question in the Appendix.

In addition, it is interesting that the majority of the strong
\ion{C}{4} systems at large velocities are in flat-spectrum QSOs as
can be seen in Figure~\ref{fig6}.  Since most of the associated
systems are strong, Foltz et al. (1986) concluded that the rest
equivalent width may be correlated with intrinsic absorption.  If this
is the case, then the excess of strong systems at high-velocities in
flat-spectrum QSOs would certainly be consistent with a population of
intrinsic absorbers.

\section{Discussion}

\subsection{Intrinsic Absorption}

A number of interesting things regarding intrinsic absorption can be
derived from the \ion{C}{4} distributions described above.  As noted,
the optically bright QSOs do indeed show an excess of absorbers and
this is true for $15,000 < v < 65,000\,{\rm km\,s^{-1}}$.  This bright
excess could be consistent with lensing of the QSOs by intervening
galaxies if a suitable lensing scenario could be found.  However,
there is a similar excess of absorbers in the flat-spectrum QSOs.
Flat-spectrum QSOs tend to be brighter than the steep-spectrum QSOs
(at least in our sample --- see the Appendix), and one might be
tempted to conclude that the flat-spectrum excess is merely a residual
of the bright excess.  We have shown in the bottom panel of
Figure~\ref{fig4} that the flat excess still persists even when we
normalize the $M_V$ distributions of the flat and steep samples, so it
seems unlikely that the flat-spectrum excess is a residual of the
bright excess.  In any case, it is difficult for gravitational lensing
by diffuse intervening galaxies to significantly alter the observed
radio spectral index of QSOs and we are inclined to believe that there
is a real excess of absorbers in flat-spectrum QSOs that is {\em not}
due to lensing of the type noted.  On the other hand, an increase in
the optical brightness could reasonably be correlated with a flatter
radio spectral index, since the optical light from the core is
probably beamed along with the radio flux from the core.  Therefore,
the bright excess {\em could} reasonably be a residual of the
flat-spectrum excess, which is probably just an orientation effect.

If a significant fraction of the \ion{C}{4} absorbers are indeed
intrinsic to the QSO environment, then it would be desirable to
explain the above observations with a single model.  If we consider
just the RLQSOs, and we assume that the spectral index correlates
with the orientation of the QSO to our line of sight, we may be
able to explain the bright/faint differences as being due to
relativisitic beaming.  Specifically, flat-spectrum QSOs may be
expected to be brighter than steep-spectrum QSOs due to beaming of
the optical continuum along with the radio continuum (\cite{bw85}).

If the clouds causing the absorption are not moving radially with
respect to the QSO engine, then orientation (and therefore spectral
index) effects might be expected in the velocity distribution of the
absorbers.  A possible model is one in which the clouds are
constrained to move along field lines which are perpendicular to the
disk.  Figure~\ref{fig7} shows a model (kindly provided by Arieh
K\"{o}nigl and John Kartje) that depicts the plane of the disk and the
magnetic field lines as being perpendicular (see also Fig. 13 in
\cite{kk94}).  In Figure~\ref{fig7}, a line of sight to the central
engine that passes through point 2 would generally show a flatter
spectrum than a line of sight passing through point 3.  However, this
is not to say that the locations of points 2 and 3 are indicative of
lines of sight that are inherently flat and steep, respectively.  If
the jet axis is perpendicular to the disk, then we might expect to see
fewer high-velocity clouds along the line of sight to steep-spectrum
QSOs as compared to flat-spectrum QSOs, since the velocity vector of
the clouds would be seen in projection towards steep-spectrum objects.
Specifically, in steep-spectrum QSOs we might expect to see a pileup
of \ion{C}{4} near zero velocity if the velocity vectors of the clouds
are generally perpendicular to our sight line, or possibly even an
evacuated velocity region as a result of clouds moving out of the line
of sight.

Another possibility is that the clouds seen in steep-spectrum objects
are pushed to higher velocities than clouds in flat-spectrum objects,
which might explain the apparent dearth of clouds (in velocity space)
towards steep-spectrum QSOs.  That is, there might be the same number
of clouds towards both types of QSOs, but for some reason the clouds
in steep-spectrum objects are more widely distributed in velocity
space.  Some of these clouds may even be pushed to velocities that
would place them in the Lyman $\alpha$ forest, which is beyond the
velocity limit of this analysis.

If there were a similar sort of orientation effect for the RQQSOs,
then we might be able to explain why the bright/faint difference is
stronger than the steep/flat difference.  That is, if we had some
measure of the orientation of RQQSOs, we might also find a
relationship between $d{\cal N}/d\beta$ and orientation angle.  Then,
if for some reason the RQQSOs also showed an optical brightening with
increasing angle from the disk, we might expect that the difference
between the bright and faint samples in both RL and RQQSOs would be
stronger than for the steep versus flat samples in RLQSOs.  If this is
the case, $M_V$ may serve as a surrogate measure of orientation in the
RQ population.

In light of the above discussion, it is interesting to note that the
small excess of absorbers in RQQSOs over RLQSOs might also be
explained as an orientation effect.  If the very flattest of the RL
QSOs are classified as BL Lacs (Fig.~\ref{fig7}, sight line 1) rather
than as QSOs, then this effectively ``steepens'' the average RL
spectral index (or orientation) with respect to the RQ population
(assuming that RL and RQQSOs are distinct populations with otherwise
similar attributes).  Since we also remove BALQSOs from our analysis,
there might be a similar biasing of the orientation in RQQSOs.
Specifically, if BALQSOs are just normal RQQSOs, but oriented such
that the line of sight skims the surface of a surrounding torus that
is coplanar with the accretion disk (Fig.~\ref{fig7}, sight line 4),
then the removal of these QSOs would ``flatten'' the average spectral
index of the RQQSOs in our sample.  This would further amplify the
difference between our RL and RQ population and the difference between
the two might be explained as a difference in the average orientation
angle of the two types of QSOs.  On the other hand if BALs are
observed in an orientation that is more perpendicular to the plane of
the disk, then this would not help explain the differences between our
RL and RQ samples.

Finally, if we make the naive assumption that all the absorbers at
velocities greater than $5000\,{\rm km\,s^{-1}}$ in steep-spectrum
QSOs are due to intervening galaxies and that the flat-spectrum excess
is entirely the result of contamination by intrinsic absorption, then
we can make a rough estimate of the fraction of intrinsic \ion{C}{4}
absorption.  Using the data from our normalized $250\,{\rm
km\,s^{-1}}$ sample, this method yields $36\%$ as an estimate of the
contamination of \ion{C}{4} systems by absorbers that are intrinsic to
the QSO.  If true, then this certainly has significant consequences
for future (and past) studies of QSO absorption line systems.
Furthermore, preliminary analysis of our core-to-lobe ratios would
seem to indicate a trend between core-to-lobe ratio and \ion{C}{4}
absorber velocity distribution.  We expect that a full reanalysis
using core-to-lobe ratios will reveal that we have mis-classified more
flat-spectrum QSOs than steep-spectrum QSOs, which may serve to
increase the dichotomy in the absorber velocity distribution between
the two populations.  If this preliminary observation can be confirmed
with a much larger sample, then the fraction of \ion{C}{4} that is
intrinsic may be significantly larger.

\subsection{Gravitational Lensing}

Vanden Berk et al. (1996) showed that there is an excess of \ion{C}{4}
absorbers seen along the line of sight to ``bright'' QSOs and argued
that this excess might be the result of each absorber causing a small
amount of magnification due to gravitational lensing.  However, they
could not rule out the possibility that the effect might be caused, at
least in part, by intrinsic absorption.  Holz \& Wald (1998)
conducted a study to determine if the effect could indeed be caused by
gravitational lensing.  They found that even for the most favorable
conditions, lensing could only account for a brightening of the QSO by
about 0.08 mag per absorber, which is not enough to produce the
observed effect, though it is in the right direction.  Herein we have
considered the alternative possibility, which is that the effect is
caused by intrinsic absorption.  We find that the existence of a
component of \ion{C}{4} absorbers that are intrinsic to the QSO is
indeed consistent with the excess of absorbers in bright QSOs as seen
by Vanden Berk et al. (1996), and also with the observations presented
herein.  However, we cannot rule out gravitational lensing (or some
other effect) without more data and further analysis.

One caveat is that since gravitational lensing is achromatic, we have
naively assumed that lensing by intervening galaxies would affect our
6 and $20\,{\rm cm}$ flux densities equally, such that lensing could
not influence the radio spectral indices of the QSOs.  However, it may
be possible for lensing to cause a flattening of the spectral index
due to the fact that lensing would have a greater effect on light from
the core than light from the lobes.  For both the core and the lobes
considered separately, there will be no relative change in the 6 to
$20\,{\rm cm}$ flux densities due to lensing, but if the core is
affected more by lensing then this changes the fraction of the light
in the beam that is coming from the core.  Since the core generally
has a flatter spectrum than the lobes, we might expect to see a
flattening of the spectrum as a result of gravitational lensing.
Therefore, an excess of absorbers in flat-spectrum QSOs could be
indicative of gravitational lensing.  This would certainly be in
agreement with the fact that our flat-spectrum QSOs are brighter than
the steep-spectrum sources by about $0.75\,{\rm mag}$.  However, it
has been found that flat-spectrum sources are generally at least 1 mag
brighter than steep-spectrum sources due to relativistic beaming
(\cite{bw85}).  Therefore, unless flat-spectrum QSOs are
preferentially lensed, there appears to be no need to invoke
gravitational lensing to explain the difference in brightness between
our flat-spectrum QSOs and our steep-spectrum QSOs.  A detailed
comparison of core-to-lobe ratios from 20\,cm FIRST maps between our
sample of QSOs and all the QSOs found in the FIRST survey may allow us
to make a more definite conclusion.

\subsection{$d{\cal N}/dz$ versus $d{\cal N}/d\beta$}

Since it has become common to assume that absorption systems with
$v>5000\,{\rm km\,s^{-1}}$ are due to intervening galaxies, it has
also become common to use $d{\cal N}/dz$ to study the evolution of
absorbers in redshift.  Whatever one's opinions are in terms of
whether it is more appropriate to study absorption systems in redshift
space or velocity space, it is worth taking a moment to consider the
differences in the two selection functions (i.e. the number of times
an absorber {\em could} have been seen at a given redshift or
velocity).  For the majority of the QSOs in our catalog, the C\,{\sc
iv} emission line is seen.  In terms of the \ion{C}{4} absorbers, this
means that the selection function has a maximum near zero velocity and
decreases smoothly with increasing velocity.  This means that the
$d{\cal N}/d\beta$ selection function is relatively flat (at least out
to the velocity of the Lyman $\alpha$ emission line, $v \approx
0.238c$).  Thus, the process of normalizing the number of absorbers
observed by the number of times an absorber could have been seen at
that velocity is unlikely to produce strong, artificial features.

On the other hand, the selection function for $d{\cal N}/dz$ is
generally {\em not} flat, since for any given data set there is
usually a peak in the redshift distribution.  In our data set, the
selection function is not very smooth as a result of the combination
of many surveys into one.  The result is that small errors in either
the observed distribution of absorbers with respect to redshift and/or
the number of times an absorber could have been detected at a given
redshift, can produce large deviations from the true value of $d{\cal
N}/dz$.  This is particularly true at the edges of the selection
function.  Although this is by no means an argument for the intrinsic
nature of \ion{C}{4} absorbers, we do feel that it is important to
point out that an analysis of the distribution of absorbers in
redshift space is much more complicated than the same analysis in
velocity space.

\subsection{Associated Absorption}

We explained that there is apparently no significant difference in
$d{\cal N}/d\beta$ (using a single $10,000\,{\rm km\,s^{-1}}$ bin) for
$v < 5000\,{\rm km\,s^{-1}}$ for any two QSO properties after having
properly removed the effects of other QSO properties.  Although this
is seemingly in contrast with other work, we emphasize that the
velocity scale studied here is quite different from that studied by
Foltz et al. (1986), whose work we verify if we use $1500\,{\rm
km\,s^{-1}}$ bins.  In addition, the similarity between the steep and
flat $d{\cal N}/d\beta$ levels could be an artifact of the fact that
our steep/flat dividing line is unphysically motivated.  If so, there
might be a real difference in the associated population between steep-
and flat-spectrum QSOs, but it is masked because we allow
steep-spectrum QSOs to creep into our flat-spectrum sample.  We hope
to revisit this problem using radio core-to-lobe ratios as a measure
of orientation in place of radio spectral indices.

However, if there really is no large scale difference in the overall
level of $d{\cal N}/d\beta$ for absorbers with velocities less than
$5000\,{\rm km\,s^{-1}}$, then this symmetry is consistent with the
associated absorber population being composed of virialized material
(such as might be caused by a cluster surrounding the QSO), since the
orientation of the QSO should not have any significant impact on the
distribution of a virialized population.  Again, this should be
considered in more detail, but it is beyond the scope of this paper.

\subsection{Line-locking}

This study was initially motivated in part as a search for
line-locking in our absorption line catalog.  Despite that fact that
we have much more data than was considered in the Burbidge \& Burbidge
(1975) study, we still find it difficult to do a proper study of
line-locking.  We estimate that 5 times as many data are necessary to
properly study emission-absorption line-locking and 60 times as many
data are required for a detailed study of absorption-absorption
line-locking.  The data from the Sloan Digital Sky Survey should
certainly provide the quantity and quality of data needed for such
studies.  However, even in the current data set there are certainly
hints of line-locking features (e.g., peaks near 0.1c) and we hope to
pursue a full analysis at a later date.  For now, we have chosen to
concern ourselves with the large scale velocity distribution of the
absorbers for which we have much better statistics.

\section{Conclusion}

We have analyzed the velocity distribution of \ion{C}{4} absorption
line systems in terms of intrinsic QSO properties for the largest
compiled sample of \ion{C}{4} systems.  This data set is a compendium
of the data in the literature and as such is indisputably
heterogeneous; however, great care has been taken to ensure that our
results are not dependent upon any known biases in our catalog.

Our conclusions regarding the distribution of \ion{C}{4} absorbers at
velocities exceeding $5000\,{\rm km\,s^{-1}}$ can be summarized as
follows:

1. There appears to be no significant change of the \ion{C}{4}
velocity distribution with respect to redshift, which is consistent
with $d{\cal N}/dz$ being independent of redshift.

2. Optically bright QSOs show an excess of \ion{C}{4} absorption
from $15,000\,{\rm km\,s^{-1}}$ to $65,000\,{\rm km\,s^{-1}}$
(relative to the emission-lines), with a peak near $\beta = 0.1c$.

3. We find an excess of absorption systems in radio-quiet QSOs over
radio-loud QSOs out to velocities of $30,000\,{\rm km\,s^{-1}}$, which
is curiously near the average terminal velocity of broad absorption
lines in BALQSOs --- despite the fact that we have excluded BALQSOs
from our analysis.

4. There is an excess of \ion{C}{4} absorption line systems in
flat-spectrum, radio-loud QSOs, with an accompanying dearth in
steep-spectrum radio-loud QSOs.  This excess is strongest from
$5000\,{\rm km\,s^{-1}}$ to $25,000\,{\rm km\,s^{-1}}$, but may extend
all the way to the Lyman $\alpha$ forest at $\beta = 0.238$.
Preliminary analysis of $20\,{\rm cm}$ data from the FIRST VLA Survey
corroborates the assumption that spectral index is roughly correlated
with QSO orientation.

5. If the observed difference in the flat-spectrum distribution as
compared to the steep-spectrum distribution is due to
orientation-dependent intrinsic absorption then we conclude that a
significant fraction ($\sim 36\%$) of \ion{C}{4} may actually be
intrinsic to the QSOs and not due to intervening galaxies.

6. The use of core-to-lobe ratios instead of spectral indices is
likely to allow for a better determination of the fraction of
intrinsic \ion{C}{4} absorption.  The few data we have now support the
observed differences in steep versus flat-spectrum QSOs and may
indicate an even larger fraction of intrinsic absorption.

Although it is possible that part of these observed effects might be
the result of gravitational lensing of QSOs by galaxies producing
absorption line signatures, we find that the observed results can be
explained without having to invoke the lensing hypothesis.  If the
velocity distribution of these clouds is not spherically symmetric,
then the observed results may not be unreasonable.  In particular, the
excess of absorbers in flat-spectrum, optically bright and radio-quiet
QSOs may be a results of relativistic beaming and orientation effects.

Regardless of what causes the observed differences in the \ion{C}{4}
distributions, it is noteworthy that there are any differences at all.
Serious consideration must be given to the consequences for studies of
QSOALS if these results can be confirmed with more homogeneous data
sets.  In particular, we stress the need for QSO absorption line
surveys with more uniform distributions of radio properties and
fainter magnitude limits.  Such surveys would help discriminate
between real effects of intrinsic absorbers as opposed to effects
resulting from using magnitude-limited samples with biased radio
properties.


\acknowledgments

A large number of people deserve our thanks for contributions they
have made to this work, including: Bob Becker for providing FIRST 20
cm contour plots and for discussions regarding the radio analysis,
Gary Sowinski for help classifying the BAL properties of our QSOs,
Jean Quashnock for discussions regarding statistical analysis, Arieh
K\"{o}nigl and John Kartje for discussions regarding terminal
velocities of BAL clouds and BAL models, Chris Mallouris and Damian
Bruni for their efforts in updating the catalog, Arlin Crotts for
discussions regarding the binning of data with different velocity
resolution, and an anonymous referee for suggestions that helped
clarify the paper.  This research has made use of the NASA/IPAC
Extragalactic Database (NED) which is operated by the Jet Propulsion
Laboratory, California Institute of Technology, under contract with
the National Aeronautics and Space Administration.

\appendix

\section{Cataloging Radio Data}

The radio data in the catalog has been compiled as follows.  First we
searched for matches between QSOs in our absorption line catalog and
the catalog of strong $1.4\,{\rm GHz}$ sources produced by White \&
Becker (1992) from the Green Bank $1.4\,{\rm GHz}$ Northern Sky Survey
(\cite{cb85}; \cite{cb86}).  Any QSO within $160\arcsec$ of a radio
source was taken as a match; matches with separations much larger than
$30\arcsec$ were confirmed with NED\footnote[4]{The NASA/IPAC
Extragalactic Database (NED) is operated by the Jet Propulsion
Laboratory, California Institute of Technology, under contract with
the National Aeronautics and Space Administration.}.  This radio
catalog has the useful property that the authors have cross-correlated
it with their $4.85\,{\rm GHz}$ catalog (\cite{bwe91}) made from the
Green Bank $4.85\,{\rm GHz}$ Northern Sky Survey (\cite{cbs89}), so
that many of the sources have not only $20\,{\rm cm}$ flux densities,
but also $6\,{\rm cm}$ flux densities and therefore a $6$ to $20\,{\rm
cm}$ radio spectral index.  White \& Becker (1992) estimate that a
comparison of sources in the $1.4\,{\rm GHz}$ catalog with those in
the $4.85\,{\rm GHz}$ catalog for sources separated by less than
$300\arcsec$ will result in $\approx 1.4\%$ of such matches being
chance coincidences.  It is expected that the matches between our
optical positions and their radio positions using a radius of
$160\arcsec$ (the $90\%$ confidence limit on the radio positions) will
result in very few spurious matches.

We then updated the $6\,{\rm cm}$ fluxes found above with those from
the GB6 Catalog (\cite{gsd+96}); nondetections within the area covered
by the GB6 catalog were assigned upper limits from this catalog.  The
$20\,{\rm cm}$ data was then supplemented from both the NVSS
(\cite{ccg+98}) and FIRST (\cite{bwh95}) $20\,{\rm cm}$ VLA surveys.
Specifically, we used the integrated flux densities from the NVSS
survey (which covers the whole sky north of -40 degrees in
declination) as this quantity should be close to the peak flux density
as measured at Green Bank.  Comparison of the two data sets bears out
this hypothesis.  Nondetections within the NVSS boundary were assigned
upper limits of $2.5\,{\rm mJy}$.  We then searched the FIRST catalog
to improve our flux density limits wherever possible, since it has a
smaller flux density limit.  However, detections from the FIRST
catalog have not been used, since the Green Bank data is at much lower
resolution than the FIRST data.

\section{RL versus RQ Determination}

Historically the division between radio-quiet and radio-loud QSOs has
been determined by either the ratio of the radio flux at $6\,{\rm cm}$
to the optical flux at $2500\,{\rm \AA}$, or by the $6\,{\rm cm}$
radio luminosity alone.  In attempting to classify the QSOs in the
catalog we have encountered a number of problems with using these
methods and/or wavelengths.  As a result we define a slightly
different set of criteria which are discussed herein.

Taking the mean QSO redshift, ${\bar{z}_{em}} \simeq 2.0$, and the
mean B magnitude, ${\bar{{\rm B}}} \simeq 18.1$, (taking B-V = 0.3) of our
catalog, along with a radio spectral index of $\alpha_{rad} \simeq
-0.5$ and an optical spectral index of $\alpha_{opt} \simeq -1.0$,
then using the traditional formula for the ratio of radio and optical
flux densities:
\begin{equation}
\log R^* = \log f(5\,{\rm GHz}) - \log f(2500\,{\rm \AA}),
\end{equation}
where 
\begin{equation}
\log f(5\,{\rm GHz}) = -29.0 + \log S_{\nu} + \alpha_{rad} \log
\left(5/\nu\right) - (1 + \alpha_{rad}) \log (1 + z_{em}),
\end{equation}
and
\begin{equation}
\log f(2500\,{\rm \AA}) = -22.62 - 0.4{\rm B},
\end{equation}
(\cite{smw+92}) then we find that a $6\,{\rm cm}$ flux density as low as
2.43 mJy will give a value of $\log R^*$ equal to 1.0 --- the value
often taken as the dividing line between RL and RQ.

This would not be a problem except that the largest, most uniform
$6\,{\rm cm}$ catalog to date, the GB6 catalog from Green Bank
(\cite{gsd+96}), has a flux limit of $18\,{\rm mJy}$ {\em at best}.
Thus, using this catalog alone would result in few (if any) RQQSOs.
However, this is not the case at $20\,{\rm cm}$.  Between the FIRST
and NVSS VLA surveys at $20\,{\rm cm}$, most of the sky has been
covered to flux density limits of at least $2.5\,{\rm mJy}$.
Therefore we chose to use $20\,{\rm cm}$ luminosity as our measure of
radio strength rather than $6\,{\rm cm}$ luminosity with the
realization that the two wavelengths are likely to sample somewhat
different properties.  Also, we chose to use the radio luminosity
rather than a ratio of radio to optical flux to avoid potential biases
caused by partial absorption of the optical flux without an
accompanying loss of radio flux.

The equation for computing the radio flux density at $20\,{\rm cm}$ in
mJy (in the rest frame) is
\begin{equation}
\log f(1.4\,{\rm GHz}) = \log S_{\nu} + \alpha_{rad} \log \left(1.4/\nu\right) - (1 +
\alpha_{rad}) \log (1 + z_{em}).
\end{equation}
Where possible we use the measured flux density at $20\,{\rm cm}$ and
the measured $6$ to $20\,{\rm cm}$ spectral index.  However, if the
$20\,{\rm cm}$ flux density was not available but the $6\,{\rm cm}$
flux density was, then we used the $6\,{\rm cm}$ value with the median
radio spectral index ($\alpha_{rad} \simeq -0.5$).

The uncertainty in $\log f(1.4\,{\rm GHz})$ due to errors in the
spectral index and errors in the radio measurements is
\begin{equation}
\sigma_{\log f(1.4\,{\rm GHz})} = \sqrt{\frac{\sigma_{S_{\nu}}^2}{S_{\nu}^2} + \sigma_{\alpha_{rad}}^2 \left(\log \left(1.4/4.85\right)\right)^2 + \sigma_{\alpha_{rad}}^2 \left(\log (1+z)\right)^2}. 
\end{equation}
At the median redshift of the catalog, $z_{em} \simeq 2.0$ and
assuming 20\% error in the radio measurements, and an error in
$\alpha_{rad}$ of $\sigma_{\alpha_{rad}} \approx 0.5$, the error is
$\sigma_{\log f(1.4\,{\rm GHz})} \simeq 0.4$.

The intrinsic, monochromatic luminosity $L$ is then given by
\begin{equation}
\log L\,({\rm ergs/s/Hz}) = \log f({\rm mJy}) - 26.0 + \log\left(4\pi\right) + 2 \log
D({\rm cm}).
\end{equation}
For the computation of luminosity distances from redshifts, we have
taken $H_o = 65\,{\rm km/s/Mpc}$ and $q_o = 0.5$, so that
\begin{equation}
D = \frac{2c}{H_o}\left[1+z - (1+z)^{1/2}\right].
\end{equation}
The error in the last term of the luminosity equation is then given by
\begin{equation}
\left[\left(\frac{2}{D}\right)^2\left(\sigma_{H_o}^2 \frac{D^2}{H_o^2}
+ \sigma_z^2 \left[\frac{2c}{H_o}\left(1 - 0.5\left(1+z\right)^{-1/2}\right)\right]^2\right)\right]^{1/2},
\end{equation}
which comes out to about 0.20 with the assumption of a 10\% error in
the Hubble constant, $\sigma_z = 0.001$, and $z_{em} = 2$.

We use these errors to minimize confusion of RQ and RL sources at the
weak detection limit.  If one were to take a specific value of $\log
L_{rad}$ and say that QSOs above some value are loud and those below
that same value are quiet, then given the magnitude of the error
calculated above, one can see that a good number of QSOs may be
misclassified.  However, one can choose instead to have {\em two}
dividing lines such that QSOs below a certain value are quiet and QSOs
above another value are loud, whereas QSOs lying between are either
moderate in radio strength or are displaced from their proper
classification by errors.  If we set these two lines such that they
are separated by at least $2\sigma$, then we can say (to 95\%
confidence) that we have not classified anything as RL that is really
RQ and vice versa.

An examination of the distribution of radio luminosities with respect
to absolute optical magnitude (Fig.~\ref{fig8}) has led us to
conclude that the easiest way to split our sample of QSOs into RQ and
RL populations is to use a straight cut in radio luminosity.  By
visual inspection of a histogram of the data, we find that the minimum
between the two distributions is about $\log L_{rad}\,({\rm
ergs/s/Hz}) = 33.25$ (at $20\,{\rm cm}$), which is comparable to the
$6\,{\rm cm}$ value of $\log L_{rad}\,({\rm W/Hz}) \simeq 26$ (modulo
the choice of units), which has been used by previous authors (e.g.,
Stocke et al. 1992).  We have shown above that the error in the log of
the radio flux density is on the order of 0.4.  If we assume that
variability and conversion to luminosity account for an additional
error of order 0.2, then we find that the total error in the log of
the radio luminosity, $\sigma_{\log L_{rad}}$, is about 0.45.  Thus
radio-quiet QSOs in our sample have detections or upper limits below
$32.80\, {\rm ergs/s/Hz}$, whereas radio-loud QSOs are those with
detections above $33.70\, {\rm ergs/s/Hz}$, such that the two
populations are separated by $2\sigma$ (Fig.~\ref{fig8}).  Therefore
we can state to 95\% confidence that no quiet object has been
classified as loud and vice versa.  QSOs falling between the two
dividing lines are classified as radio-moderate and will be treated
separately in our analysis of absorption properties.  Upper limits
falling in the radio-loud regime (there are only four such points) will
also be treated separately, since it is quite possible that better
radio data will show that these are actually radio-quiet.

There is at least one potential problem with this division between
loud and quiet.  The dividing line between moderate and quiet is such
that there are few very high redshift radio-quiet QSOs, which is a
potential source of bias.  In fact, 27 of the upper limits at $z>3.5$
are just above our RQ cutoff line.  However, there are very few (five)
radio-loud QSOs with similar redshifts.  Essentially, this has the
effect of reducing the maximum redshift that we can probe during the
course of this study to $z_{em} \approx 3.2$.

\section{Other QSO Properties}

In addition to radio luminosity, we compare the \ion{C}{4} absorption
line distributions to the QSO emission redshift, absolute optical
magnitude, 6 to 20\,cm radio spectral index, and 20\,cm radio
morphology.

The absolute optical magnitudes are determined using an optical
spectral index $\alpha_{opt} \simeq -1.0$ and
\begin{equation}
M_V = V + 5.0 - 5.0 \log(D) + 2.5(1 + \alpha_{opt})\log(1+z_{em}),
\end{equation}
where the V magnitude and emission redshift are taken from the
literature.  Again, the distances have been calculated using $H_o =
65\,{\rm km/s/Mpc}$ and $q_o = 0.5$.  Note that the apparent magnitude
used in not always V, but may sometimes be B or R.  Unfortunately,
sources of apparent magnitudes don't always clearly indicate to which
filter the reported magnitude refers.  In these cases we assume a flat
optical spectrum and take V=B=R (though we still use $\alpha_{opt}
\simeq -1.0$ for the K-correction term).  The errors incurred by this
process are small ($<0.1\,{\rm mag}$) and are generally less than
errors due to measurement and variability and should not affect our
results.  In addition, the use of filters other than V is normally
done in a way that is appropriate to our assumption of B=V=R.  For
example, the R filter is often used to observe high redshift QSOs,
where the V band would sample the Lyman $\alpha$ forest, but the R
band samples rest wavelengths similar to that of the V band at lower
redshifts.  The median absolute magnitude for our sample is $M_V
\simeq -27.0$.

Radio spectral indices are calculated for all QSOs having detections
at both 6 and 20\,cm, where
\begin{equation}
\alpha_6^{20} = -\frac{\log\left(f_6({\rm mJy})\right) - \log\left(f_{20}({\rm mJy})\right)}{\log\left(\nu_{20}\right) - \log\left(\nu_{6}\right)}.
\end{equation}
The median spectral index for the 262 sources with both 6 and 20\,cm
measurements is $\alpha_6^{20} \simeq -0.5\pm0.5$, where the error is
the one sigma deviation for all the data (though it is not a Gaussian
distribution).

Of the 262 QSOs with measured radio spectral indices, about
one-quarter are within the area of the FIRST Survey.  R. Becker has
kindly provided us with these maps and we have classified each object
by eye as core dominated or not core dominated without any a priori
knowledge of the other QSO properties.  We find that 17 are core
dominated, whereas 21 are not, with the remaining being uncertain.
For the core dominated sources the mean spectral index is
$-0.33\pm0.45$ and the noncore sources have $-0.74\pm0.37$.  Although
the difference between the two populations is indistinguishable within
the errors, it is comforting to find that core-dominated sources
generally have a spectrum that is flatter than non-core sources as
would be expected from orientation effects in unified models
(\cite{ant93}).  We have taken $\alpha_6^{20} = -0.5$ as the dividing
line between steep and flat --- a choice which is apparently born out
by the mean core and non core values quoted above, but is otherwise
nonphysical.

In an effort to cross check our results obtained using radio spectral
indices as a measure of orientation and our preliminary classification
of FIRST sources as either core or lobe dominated, we have also
calculated radio core-to-lobe ratios for our sample of QSOs.  We use
the FIRST 5$\arcsec$ resolution survey for our core radio flux density
values and the lower resolution (45$\arcsec$) NVSS data to estimate
the total flux density from each source.  The lobe flux density is
then taken to be the difference between the NVSS total and FIRST core
flux densities.  Our values of R may be affected by variability
between the epochs of the FIRST and NVSS observations, but this effect
should be relatively small and produce no spurious trends in the data
as a source is equally likely to be in a high state either during the
FIRST or NVSS observations and hence produce slightly high or low
values of R equally as often.

The radio core-to-lobe ratio, ${\rm R}$$=$${\rm S}_{\rm core}/{\rm
S}_{\rm lobe}$, is a good indicator of relative orientation of similar
classes of active galactic nuclei (AGNs; e.g., Orr and Browne 1982).
Although lobe flux originates many kiloparsecs from the central
engine, core flux is dominated by emission from the inner jet and is
likely beamed.  Lobe emission is largely independent of source
orientation, but the strength of core flux is sensitive to the angle
that the jet axis makes with the line-of-sight becoming greatly
enhanced at small orientation angles.  All other parameters being
equal, the observed core-to-lobe ratio should therefore be largest for
those objects with jets oriented at the smallest line-of-sight angles.

We computed R only for spatially resolved sources.  Sources which are
spatially unresolved are, almost certainly, highly core dominated
sources, although it is difficult to determine actual upper limits to
the core dominance parameter.  Of the 124 quasars considered, 59 were
detected by both radio surveys and 35 of these are resolved.  All
sources were examined visually in order to assure that multiple radio
components were properly identified.

These core-to-lobe ratio calculations generally confirm the expected
trend between core-to-lobe ratios and spectral indices, though there
are a number of exceptions that are likely to be due to variability.
Of those that we classified as steep spectrum, 16 are lobe dominated
as expected, but 10 are core dominated.  For those QSOs that we
consider to be flat-spectrum, six are core dominated as expected,
whereas three are lobe dominated.  Therefore 13 of the 35 QSOs may be
misclassified in terms of their spectral indices, whereas the
remaining 22 core-to-lobe ratios are in good agreement with the
spectral indices.  A full reanalysis of the QSO orientation using
core-to-lobe ratios is beyond the scope of this paper and will be
reserved for future work.

Although we have not intended to include Broad Absorption Line QSOs
(BALQSOs) in the catalog, the large amount of attention that the BAL
phenomenon has been given in recent years (e.g., \cite{wey95b}, and
references therein) almost ensures that some of our QSOs will acquire
the BAL label.  Previous studies using data from the catalog have not
purposely included the known BALs; however, we have found that recent
lists of BAL or suspected BAL QSOs have slightly more objects in
common with our catalog than was previously thought.  As a result, we
have taken the time to classify all the QSOs in the catalog as BAL,
probable/possible BAL, CAAL (QSOs with Complex Associated Absorption
Lines), probable/possible CAAL, and not any of the above using a
version of the BALQSO catalog produced by Sowinski, Schmidt, \& Hines
(1997).

\section{Discussion of QSO Properties}

From Figure~\ref{fig8} it is obvious that there is a dichotomy between
RLQSOs and RQQSOs as has been reported numerous times in the
literature (e.g., \cite{kss+89}).  It is reassuring that we are able
to repeat the separation of the populations and that our dividing line
between the two species is close to that used by other authors, even
though we use slightly different methods.  One striking fact is that
there are about as many radio-loud QSOs in our sample as radio-quiet
QSOs, despite the fact that radio-loud QSOs are only supposed to be
about 10\% of the QSO population (\cite{sw80}).  Although it is
certainly possible that some of this is the result of mismatches
between the optical and radio identifications from our literature
search for radio measurements, it is exceedingly unlikely that this
would cause the fraction of RLQSOs to go from 10\% to over 50\%.
However, the fraction of radio-loud QSOs is seen to increase
considerably for the brightest sources (\cite{hif+95}). Thus we would
expect a fraction larger than 10\% in a population of QSOs selected as
targets for spectroscopy, since these are likely to be the brightest
QSOs.  Even so, it is not quite clear why the fraction should be near
50\%.  One possible explanation may be that those studying QSO
absorption lines have the tendency to try to observe equal numbers of
each type of QSO (at least in the aggregate of all observers) and this
is the cause of the excess of RLQSOs in our sample.  Whatever the
cause, it is seems worth noting this discrepancy.

In Figure~\ref{fig9} we show the distribution of $M_V$ versus the
emission redshift of the QSO.  The RQ population tends to be brighter
than the RL population.  In addition, there is a lack of faint objects
at high redshifts which is to be expected as a result of flux limited
surveys of QSOs.  There is also an absence of bright objects at low
redshifts, which is probably due to the small volume sampled at lower
redshifts and/or some sort of luminosity or density evolution.

We have applied the Student's $t$-test (Press et al. 1995) to see if the
observed differences are indeed significant.  Student's $t$ is
essentially the number of standard errors that the difference in two
means deviates from the null hypothesis (that there is no difference
in the mean) and the probability is that of randomly obtaining a value
of Student's $t$ that is at least as large as $t$.  We have first made
cuts at $z=3$ and $M_V=-26.0$ to correct for the apparent dearth of
faint, high redshift objects that is caused by the limiting magnitude
of the data.  The redshift difference for the bright versus faint samples
is $\Delta z = 0.342$ with a Student's $t$ of 4.83 and a probability
of $6.2 \times 10^{-6}$.  Similarly, the absolute magnitude difference
between the high and low redshift samples is -0.39 mag with a
Student's $t$ of 4.6 and probability $1.5 \times 10^{-5}$.

There are also significant differences in the radio-loud (indicated by
squares in Fig.~\ref{fig9}) and the radio-quiet (indicated by
arrows) populations.  The average redshift of the RL sample is 1.76,
whereas for the RQ sample it is 2.28.  The difference has a Student's
$t$ value of 7.14 with probability $7.6 \times 10^{-11}$.  Also, the
RL sample has an average $M_V$ of -26.68, whereas the RQ value is
-27.21; this difference has a probability of $1.49 \times 10^{-6}$ for
a Student's t of 5.07.  On the other hand, for RLQSOs we find that
there is no significant difference in the radio luminosities of the
optically bright versus faint samples, in agreement with Hooper et
al. (1995).  However, we do find a small difference in the radio
luminosities of the high $z$ versus low $z$ samples with a Student's
$t$ of $3.02$ and a probability of $3.7 \times 10^{-3}$.

The distribution of radio spectral indices is plotted against absolute
optical magnitude in Figure~\ref{fig10}.  There appears to be an
excess of faint sources that are steep-spectrum and an accompanying
dearth of bright, steep-spectrum sources.  This is unusual since
although it may be harder to find lobe dominated steep-spectrum
objects at high redshifts, there should be no such bias against
finding bright steep-spectrum objects; however, it is possible that
this effect is intrinsic to the QSOs and is not an observational bias.
On the other hand, flat-spectrum QSOs seem to be fairly evenly
distributed between bright and faint.

Statistically we find that the steep-spectrum QSOs are fainter than
the flat-spectrum QSOs by $0.73\,{\rm mag}$, which gives a Student's
$t$ of 5.57 and a probability of $4.8 \times 10^{-7}$.  The difference
in spectral index between the bright and faint samples is 0.25 with a
Student's t of 4.32 and probability $5.8 \times 10^{-5}$. The linear
correlation coefficient between the spectral index and absolute
magnitude is $-0.33$.

Finally we plot radio spectral index versus redshift in
Figure~\ref{fig11}.  Here, we find that each quadrant is populated with
similar numbers of objects except for very high redshift,
steep-spectrum objects which we expect to be underpopulated in flux
limited samples due to surface brightness effects. However, if we
consider the distribution of spectral index versus redshift and we
take only QSOs with $z<3$ to correct for this lack of lobe-dominated,
high redshift QSOs, we find no significant difference in the
redshift of the steep versus flat samples, nor in the spectral index of
the high redshift versus low redshift samples.

\clearpage

\begin{figure}
\plotone{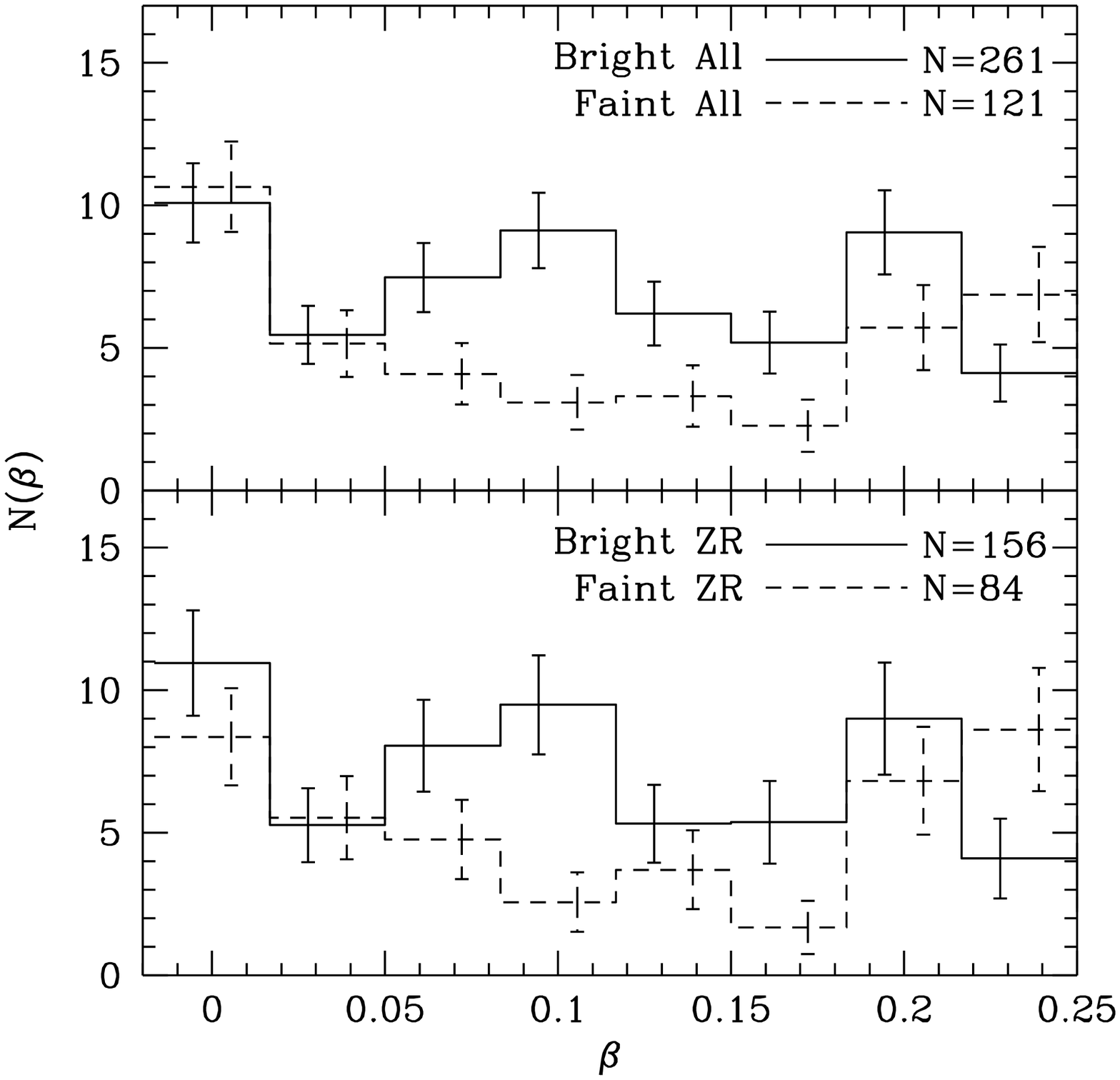}
\caption{Normalized velocity distribution of C\,{\sc iv} absorbers for
bright and faint QSOs.  {\em Top:} All absorbers graded A,B, or C and
combined within $250\,{\rm km\,s^{-1}}$.  {\em Bottom:} Same sample
but with the redshift (z) and radio luminosity (R) distributions
forced to be the same for both the bright and faint samples.  The
bright sample is given by the solid line, whereas the dashed line is
for the faint sample.  The number of absorbers in each sample is
indicated in the upper right-hand corner.\label{fig1}}
\end{figure}

\begin{figure}
\plotone{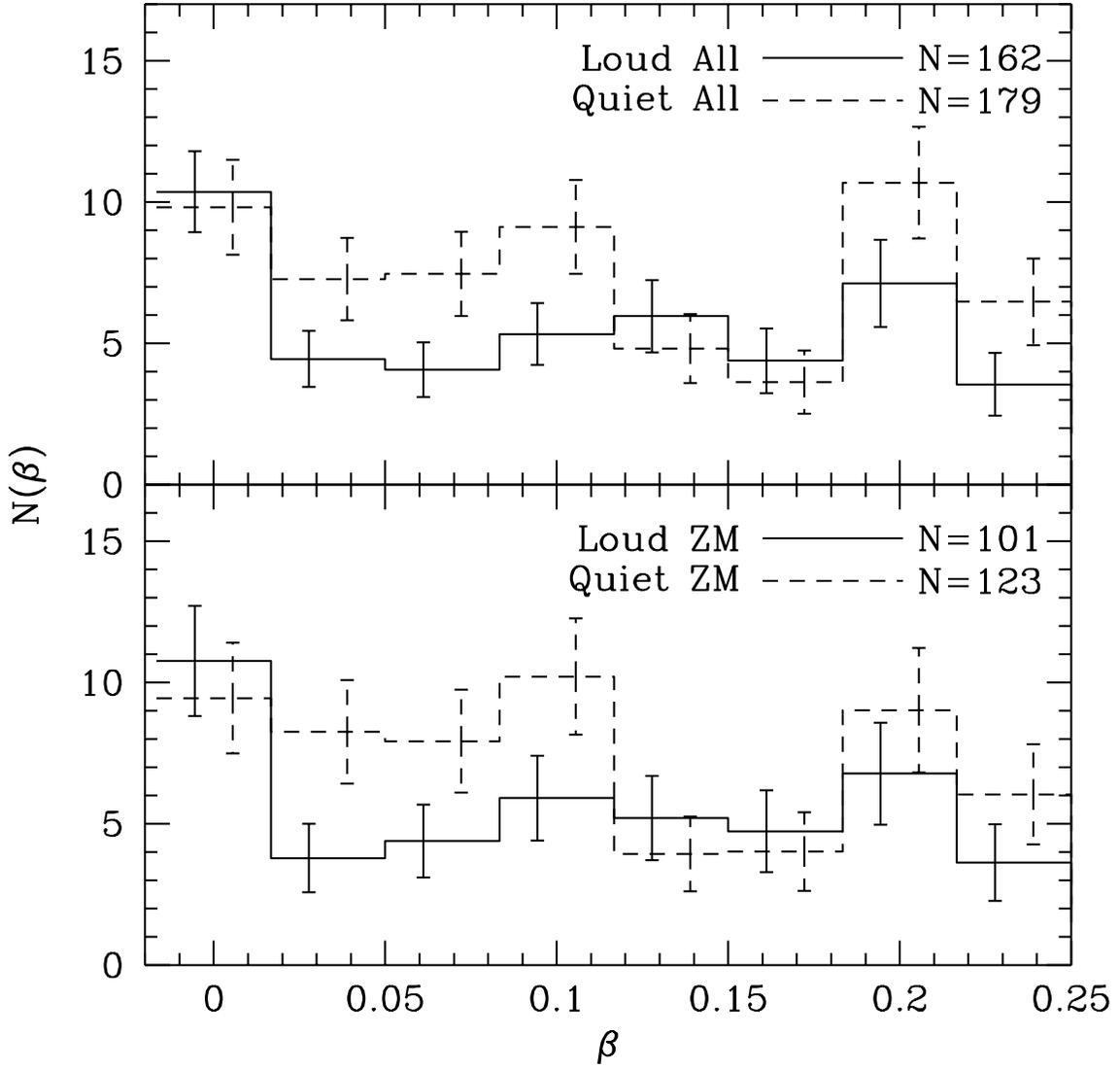}
\caption{Same as for Fig.~1 but for radio-loud and radio-quiet
QSOs. ZM in the lower panel refers to the fact that $z_{em}$ and $M_V$
are normalized for this sample. \label{fig2}}
\end{figure}

\begin{figure}
\plotone{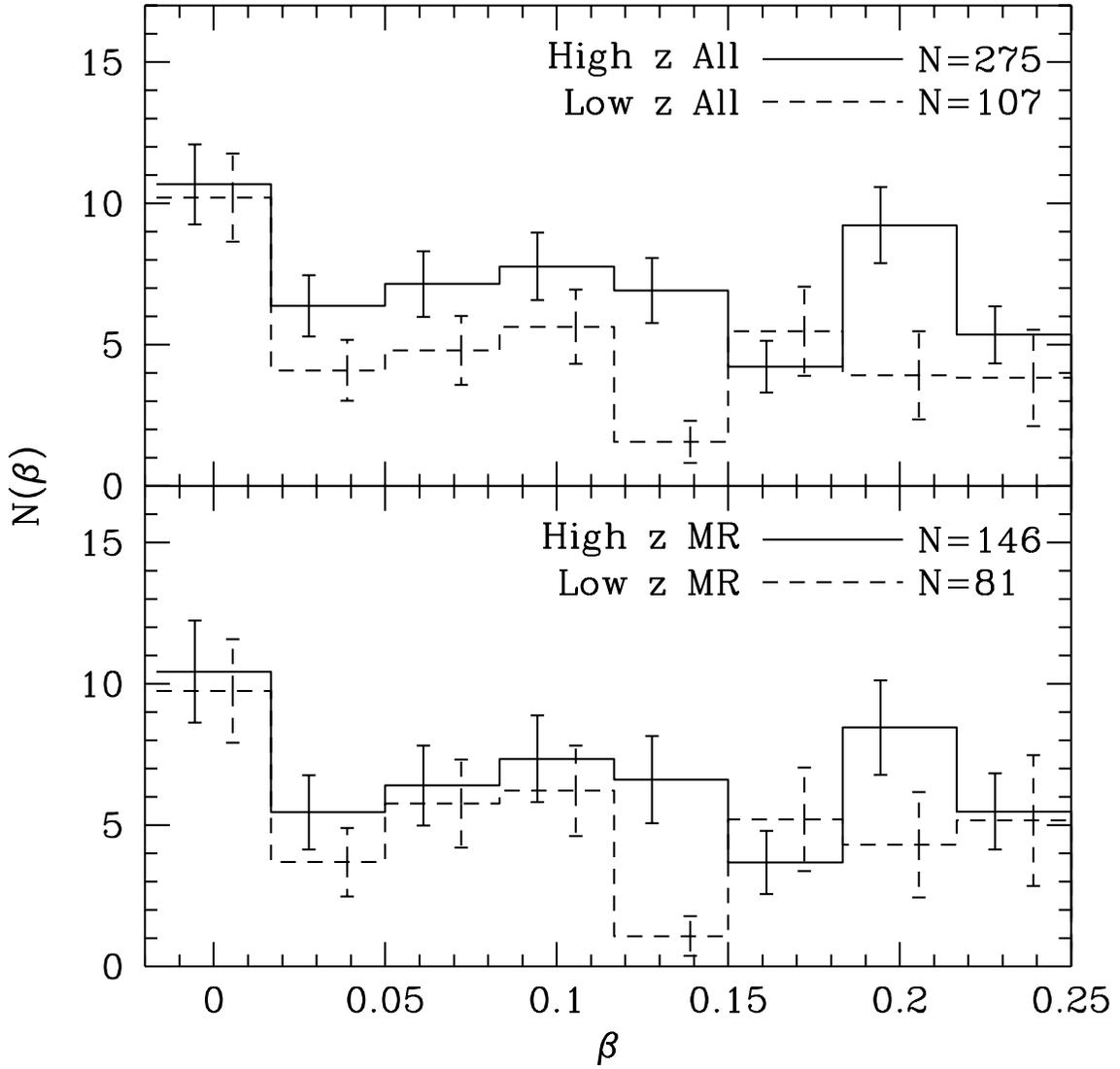}
\caption{Same as for Fig.~1 but for high-redshift and low-redshift
QSOs. The absolute visual magnitude (M) and radio luminosity (R) have been
normalized in the lower panel. \label{fig3}}
\end{figure}

\begin{figure}
\plotone{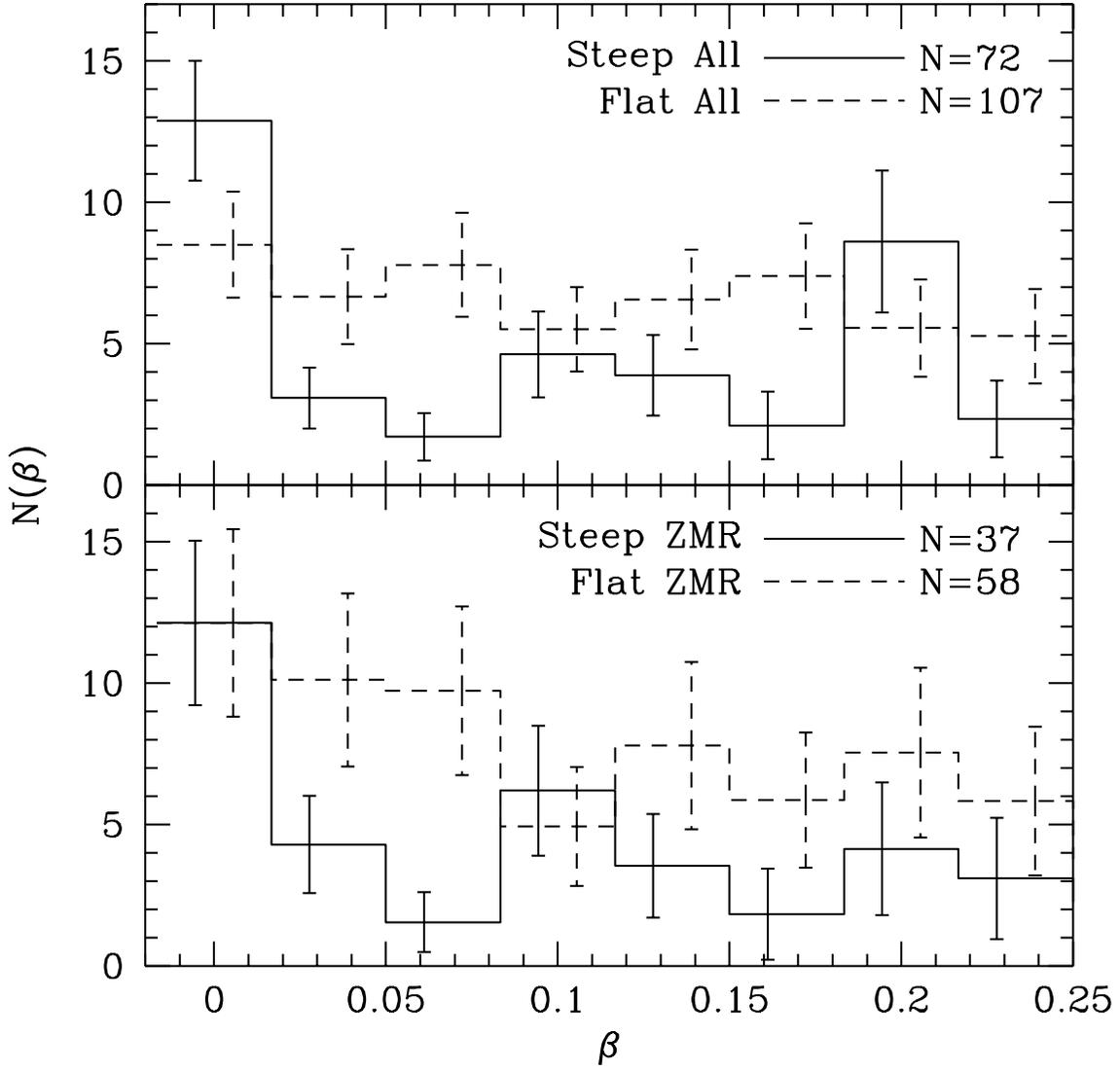}
\caption{Same as for Fig.~1 but for steep-spectrum versus flat-spectrum QSOs.  In the lower panel, $z_{em}$, $M_V$, and radio luminosity have been normalized for the two samples. \label{fig4}} 
\end{figure}

\begin{figure}
\plotone{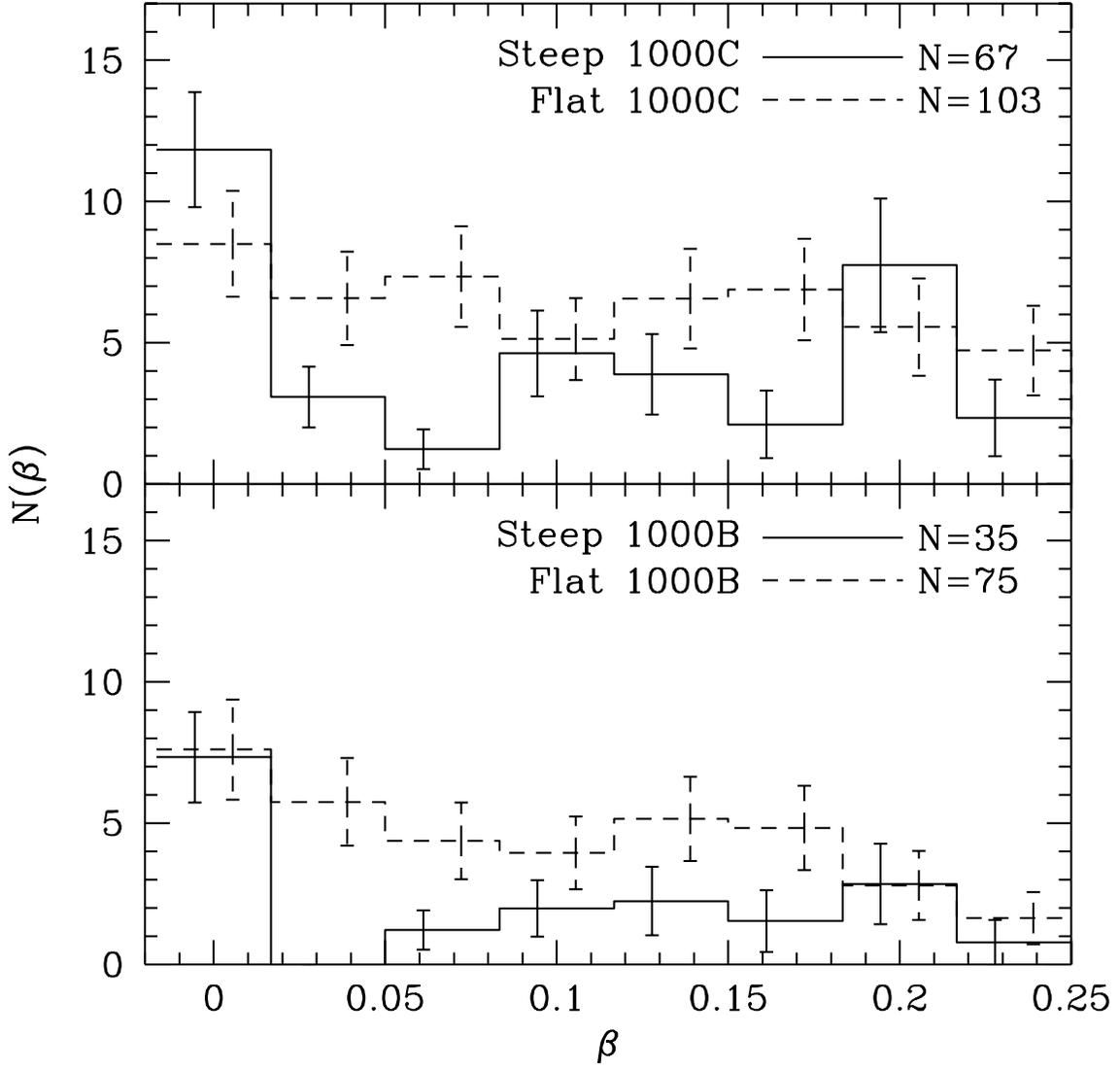}
\caption{Same as Fig.~4 but for A,B, and C graded absorbers combined within $1000\,{\rm km\,s^{-1}}$ ({\em top}) and for A and B graded absorbers combined within $1000\,{\rm km\,s^{-1}}$ ({\em bottom}). \label{fig5}} 
\end{figure}

\begin{figure}
\plotone{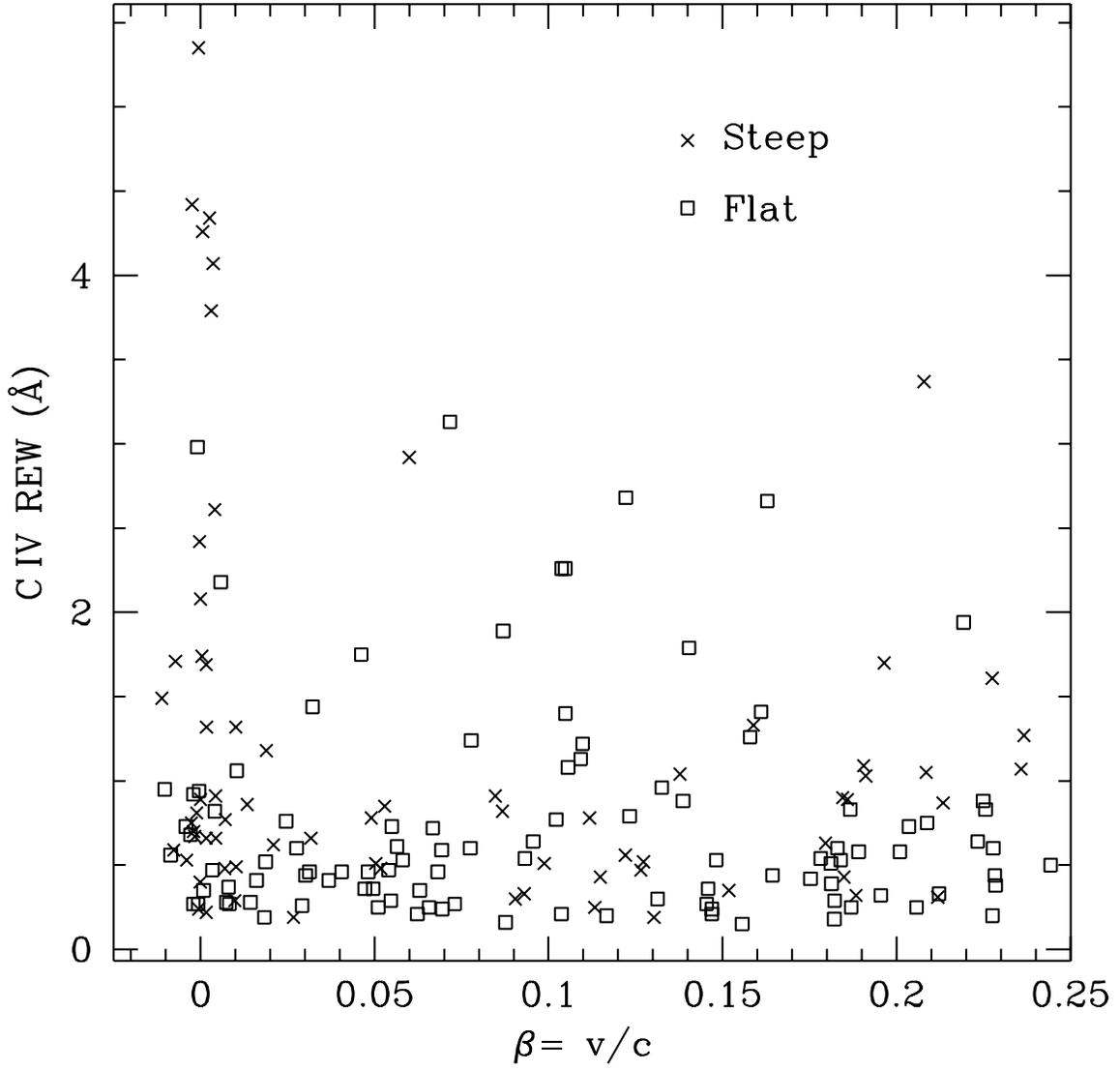}
\caption{Rest Equivalent Width (REW) versus ejection velocity
($\beta$) for both flat- and steep-spectrum QSOs. Note the excess of
strong absorbers in flat-spectrum QSOs at high
velocities. \label{fig6}}
\end{figure}

\begin{figure}
\plotone{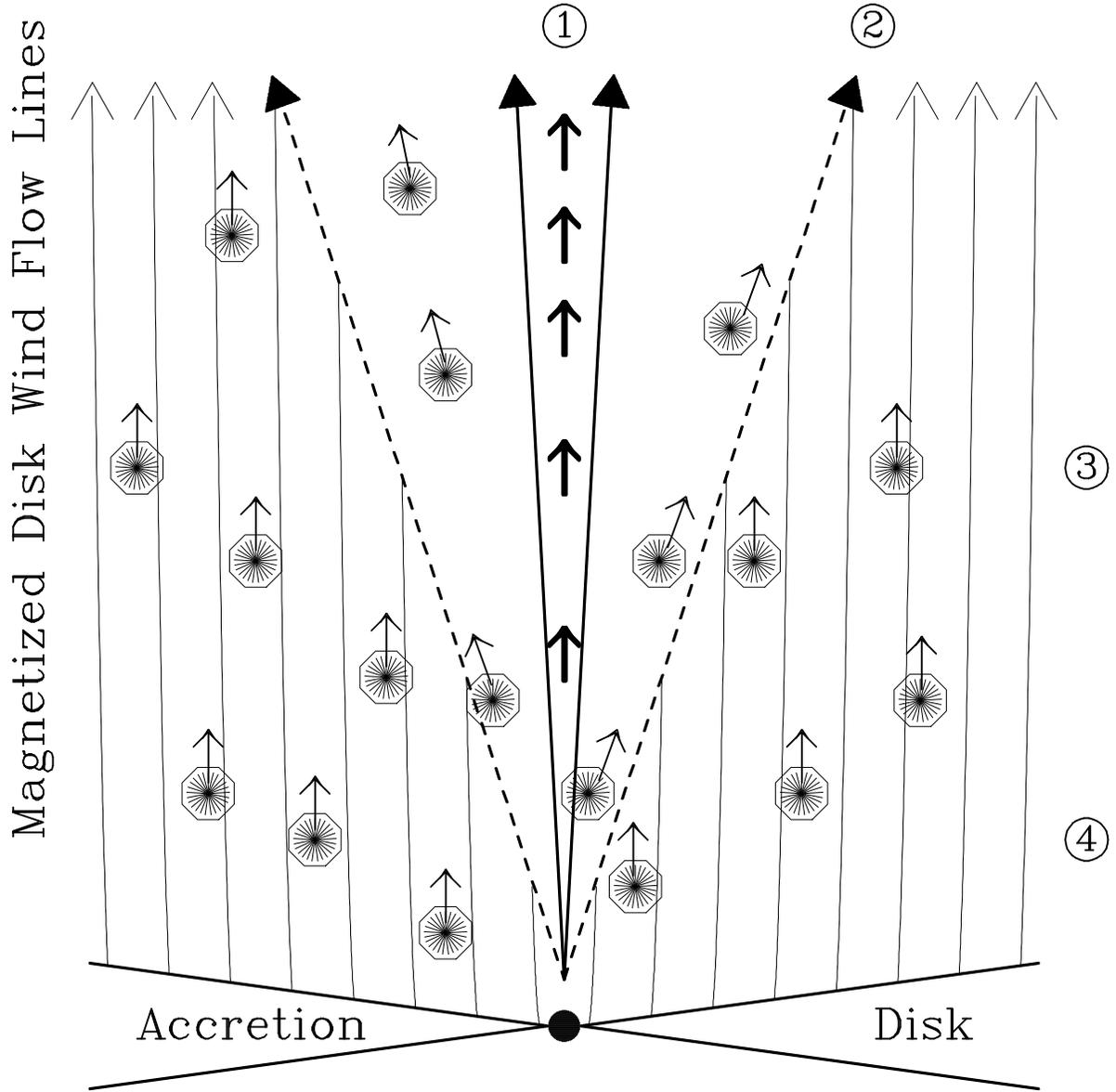}
\caption{Model of the region near the central engine of a QSO.  Here
the magnetic field lines are perpendicular to the plane of the disk.
Sight line 1 represents a BL Lac line of sight.  A line of sight
through 2 would generally have a flatter radio spectrum than a line of
sight through point 3; however, this is not to say that 2 and 3 are
lines of sight to flat- and steep-spectrum QSOs, respectively, but
rather that steep-spectrum QSOs are generally observed closer to the
plane of the disk than flat-spectrum QSOs.  If BALs are formed from
material stripped off a torus (not shown) in the plane of the
accretion disk, then a line of sight through 4 would represent
BALQSOs.  (Courtesy Arieh K\"{o}nigl and John Kartje)
\label{fig7}}
\end{figure}

\begin{figure}
\plotone{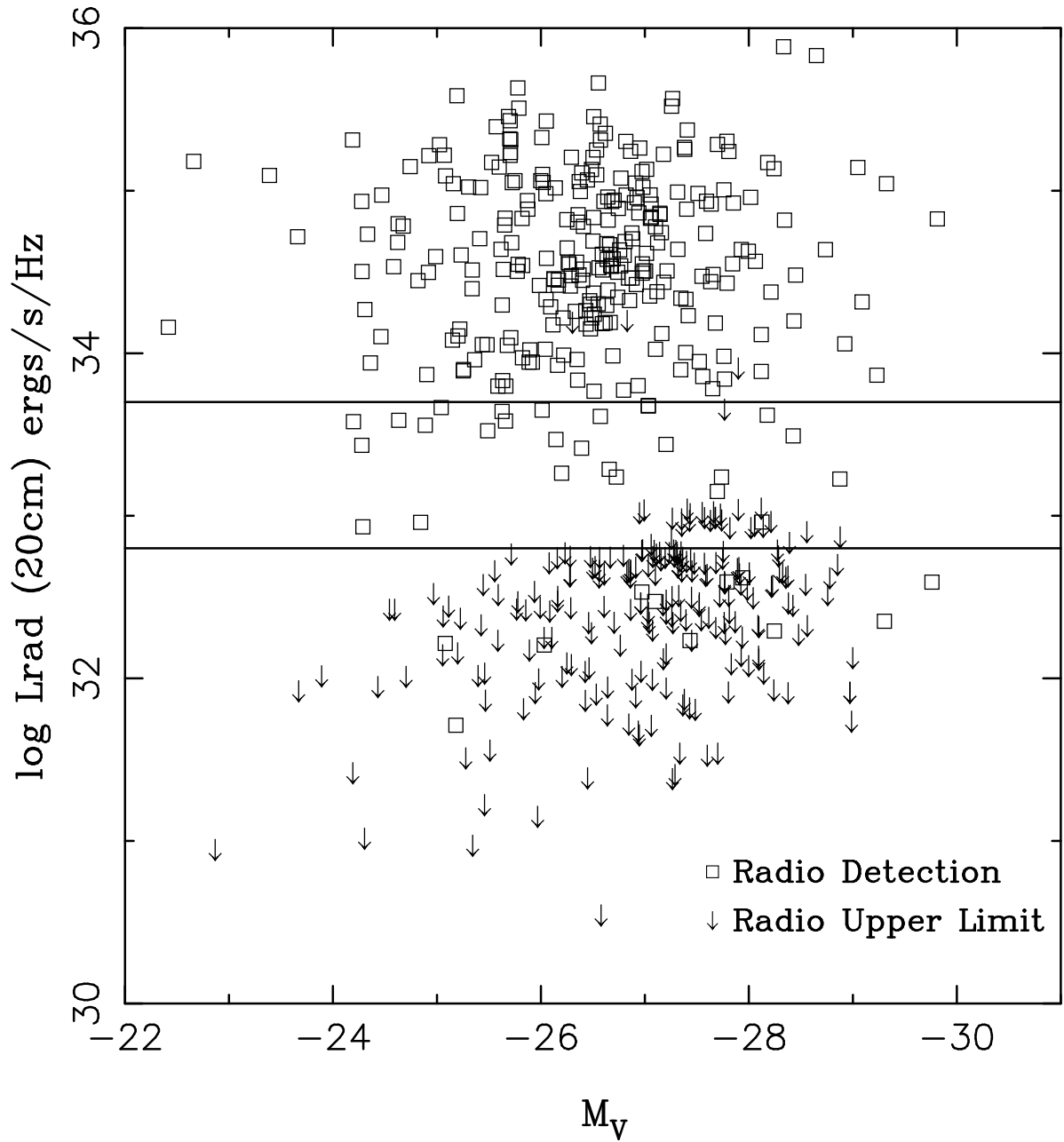}
\caption{1.4 GHz radio luminosity versus optical absolute magnitude. The two solid lines show the cuts used to determine RL versus RQ. \label{fig8}} 
\end{figure}

\begin{figure}
\plotone{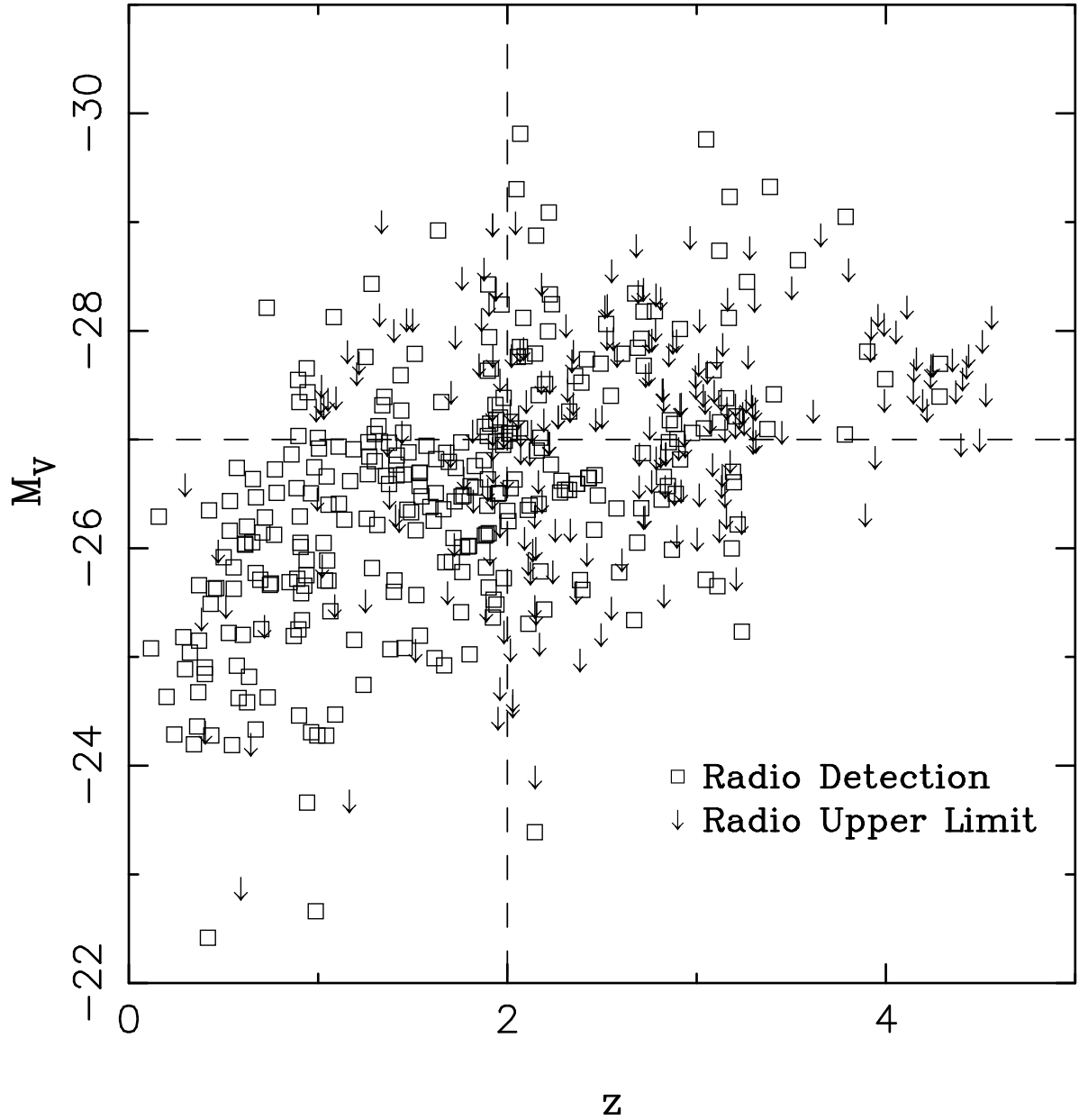}
\caption{Absolute optical magnitude versus redshift.  Arrows indicate
upper limits to the radio luminosity and are mostly RQ, whereas squares
represent detections that are largely RL.  Dashed lines indicate the
dividing lines between our samples.  Note the expected lack of high-z,
faint QSOs and bright, low-z QSOs. \label{fig9}}
\end{figure}

\begin{figure}
\plotone{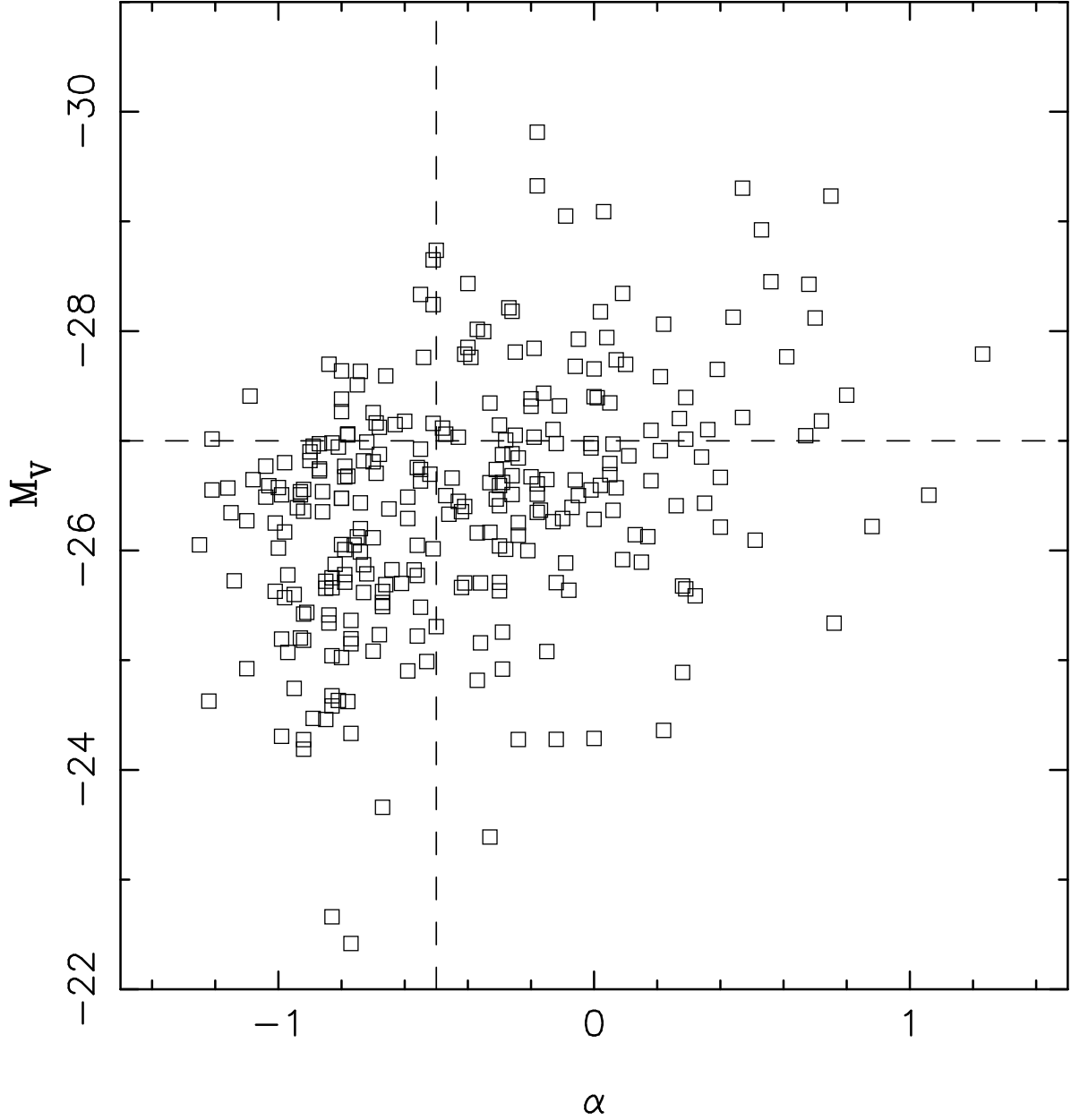}
\caption{Absolute optical magnitude versus radio spectral index.  Only
radio detections are plotted.  Dashed lines indicate the dividing
lines between QSO properties in our samples.  Note the excess of
faint, steep-spectrum objects in the lower left quadrant and the
dearth of bright, steep-spectrum objects in the upper left
quadrant. \label{fig10}}
\end{figure}

\begin{figure}
\plotone{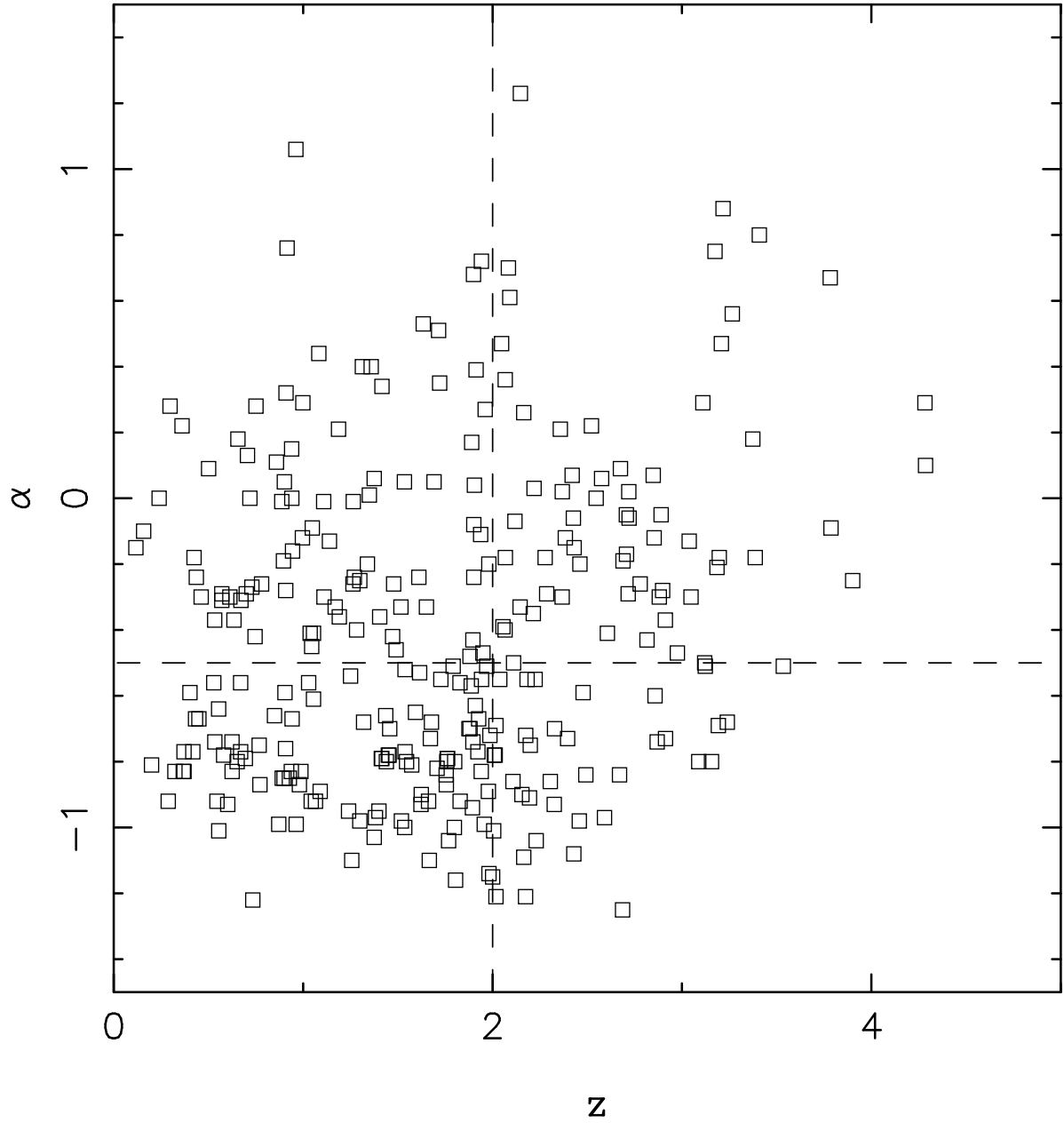}
\caption{Radio spectral index versus redshift.  Dashed lines break the plot into two our pairs of samples.  Apparent here is the expected lack of high redshift, steep-spectrum objects. \label{fig11}} 
\end{figure}

\clearpage

\begin{deluxetable}{lccccccc}
\tablecolumns{8}
\tablewidth{0pc}
\tablecaption{Mean $d{\cal N}/d\beta$ tests \label{tbl-1}}
\tablehead{
\colhead{} &
\multicolumn{3}{c}{$5000\,{\rm km\,s^{-1}}$ to $35,000\,{\rm km\,s^{-1}}$} &
\colhead{} &
\multicolumn{3}{c}{$5000\,{\rm km\,s^{-1}}$ to $65,000\,{\rm km\,s^{-1}}$} \\
\cline{2-4} \cline{6-8} \\
\colhead{Sample} &
\colhead{$\Delta d{\cal N}/d\beta$} &
\colhead{Std. Dev.} &
\colhead{Signif. ($\sigma$)} &
\colhead{} &
\colhead{$\Delta d{\cal N}/d\beta$} &
\colhead{Std. Dev.} &
\colhead{Signif. ($\sigma$)}
}
\startdata
\sidehead{Bright/Faint}
250 C & 3.236 & 0.921 & 3.511 & & 3.143 & 0.661 & 4.753 \nl
250 C ZR & 3.316 & 1.173 & 2.828 & & 2.906 & 0.839 & 3.464 \nl
1000 C & 2.694 & 0.897 & 3.002 & & 2.828 & 0.645 & 4.383 \nl
1000 B & 1.909 & 0.697 & 2.741 & & 1.989 & 0.495 & 4.014 \nl
\sidehead{Loud/Quiet}
250 C & 3.336 & 1.063 & 3.137 & & 1.945 & 0.769 & 2.530 \nl
250 C ZM & 4.096 & 1.343 & 3.049 & & 2.090 & 0.937 & 2.232 \nl
1000 C & 2.922 & 1.027 & 2.845 & & 1.730 & 0.748 & 2.315 \nl
1000 B & 1.536 & 0.796 & 1.928 & & 0.803 & 0.571 & 1.407 \nl
\sidehead{High z/Low z}
250 C & 2.254 & 0.962 & 2.344 & & 2.697 & 0.663 & 4.065 \nl
250 C MR & 1.174 & 1.179 & 0.996 & & 1.947 & 0.809 & 2.405 \nl
1000 C & 2.585 & 0.918 & 2.817 & & 2.992 & 0.635 & 4.715 \nl
1000 B & 2.011 & 0.707 & 2.844 & & 2.192 & 0.481 & 4.558 \nl
\sidehead{Steep/Flat}
250 C & 3.519 & 1.176 & 2.992 &  & 2.577 & 0.909 & 2.835 \nl
250 C ZMR & 4.246 & 1.877 & 2.262 & & 4.070 & 1.318 & 3.089 \nl
1000 C & 3.369 & 1.145 & 2.942 & & 2.562 & 0.887 & 2.888 \nl
1000 B & 3.622 & 0.896 & 4.041 & & 2.836 & 0.680 & 4.169 \nl
\enddata
\tablecomments{Given are the differences in the mean values of $d{\cal
N}/d\beta$, the standard deviation of these differences and their
significance for four different samples with respect to four sets of QSO
properties in two different velocity bins.  ZR, ZM, MR, and ZMR refere
to properties that have been normalized in a given sample, see
Figs.~\ref{fig1}--~\ref{fig5} for an explanation of the samples.}
\end{deluxetable}
\clearpage

\end{document}